\documentclass[useAMS,usenatbib]{mnras}

\usepackage[utf8]{inputenc}
\usepackage{graphicx}
\usepackage{color}
\usepackage{booktabs}
\usepackage[nolist]{acronym}
\usepackage{amssymb}
\usepackage[]{natbib}
\usepackage{float}
\hypersetup{draft}
\usepackage{hyperref}
\usepackage{amsmath,verbatim,mathtools,needspace,enumitem,etoolbox,physics,microtype}
\usepackage{epsfig}
\usepackage{url}
\usepackage{multirow}
\usepackage{array,multirow}

\newcolumntype{M}[1]{>{\centering\arraybackslash}m{#1}}
\newcolumntype{N}{@{}m{0pt}@{}}

\newcommand{\eagle}{{\sc eagle}}
\newcommand{\mobse}{{\sc mobse}}

\def\apj{ApJ}
\def\apjs{ApJS}
\def\apjl{ApJ}

\def\mnras{MNRAS}

\def\nat{Nature}
\def\araa{ARA\&A}
\def\aap{A\&A}

\def\pc{{\rm\thinspace pc}}
\def\kpc{{\rm\thinspace kpc}}

\def\Mpc{{\rm\thinspace Mpc}}

\def\Msun{\hbox{$\rm\thinspace M_{\odot}$}}

\def\yr{{\rm\thinspace yr}}

\def\Msunpc2{{\Msun\pc}^{-2}}
\def\Msunyrkpc2{{\Msun\yr^{-1}\kpc}^{-2}}

\def\magarcsec2{{\rm\thinspace mag\thinspace arcsec}^{-2}}

\begin{document}

\title[The host galaxies of compact binary mergers across cosmic time
]{Mass and star formation rate of the host galaxies of compact binary mergers across cosmic time}

\author[Artale et al.]{M. Celeste Artale$^{1}$\thanks{E-mail:Maria.Artale@uibk.ac.at, mcartale@gmail.com}, Michela Mapelli$^{1,2,3,4}$, Yann Bouffanais$^2$,
Nicola Giacobbo$^{2,3,4}$, \newauthor Mario Spera$^{1,2,3,4,5,6}$, and  Mario Pasquato$^{3,4}$\\ 
$^{1}$Institut f\"{u}r Astro- und Teilchenphysik, Universit\"{a}t Innsbruck, Technikerstrasse 25/8, 6020 Innsbruck, Austria\\
$^{2}$Physics and Astronomy Department Galileo Galilei, University of Padova, Vicolo dell'Osservatorio 3, I--35122, Padova, Italy\\
$^{3}$INAF-Osservatorio Astronomico di Padova, Vicolo dell'Osservatorio 5, I--35122, Padova, Italy\\
$^{4}$INFN-Padova, Via Marzolo 8, I--35131 Padova, Italy\\
$^{5}$Center for Interdisciplinary Exploration and Research in Astrophysics (CIERA), Evanston, IL 60208, USA \\
$^{6}$Department of Physics \& Astronomy, Northwestern University, Evanston, IL 60208, USA \\
}
\maketitle
\begin{abstract}
  We investigate the properties of the host galaxies of compact binary mergers across cosmic time, by means of population-synthesis simulations combined with galaxy catalogues from the \eagle{} suite. We analyze the merger rate per galaxy of binary neutron stars (BNSs), black hole--neutron star binaries (BHNSs) and binary black holes (BBHs) from redshift zero up to redshift six. The binary merger rate per galaxy strongly correlates with the stellar mass of the host galaxy at any redshift considered here.  This correlation is significantly steeper for BNSs than for both BHNSs and BBHs. 
  Moreover, we find that the merger rate per galaxy depends also on host galaxy's star formation rate and metallicity. We derive a robust fitting formula that relates the merger rate per galaxy with galaxy's star formation rate, stellar mass and metallicity at different redshifts. The typical masses of the host galaxies increase significantly as redshift decreases, as a consequence of the interplay between delay time distribution of compact binaries and cosmic assembly of galaxies.
Finally, we study the evolution of the merger rate density with redshift. At low redshift ($z\leq{}0.1$) early-type galaxies give a larger contribution to the merger rate density  than late-type galaxies. This trend reverts at $z\ge{}1$.
\end{abstract}


\section{Introduction}
The first and second observing runs (O1 and O2) of the LIGO-Virgo collaboration (LVC) led to the detection of ten binary black holes (BBHs) and one binary neutron star (BNS) \citep{Abbott2016a,Abbott2016b,Abbott2016c,Abbott2017,Abbott2017b,abbottGW170814,AbbottO2,AbbottO2popandrate},  marking the dawn of gravitational-wave (GW) astronomy.

The third observing run (O3) started in April 2019, and several new gravitational-wave (GW) candidate triggers have already been released. By the end of O3, we expect to have a remarkable population of several tens of BBHs, several new BNSs, and possibly the first merging black hole -- neutron star binaries (BHNSs). 

KAGRA and LIGO--India will join the detector network soon \citep[see e.g.,][]{Somiya2012,Aso2013,Abbott2016,Abbott2018}, while in about one decade from now third-generation ground-based GW detectors (Einstein Telescope and Cosmic Explorer)  will lead to a dramatic improvement of sensitivity in GW searches and will be able to catch BBH mergers up to redshift $z\ge{}10$ (see e.g. \citealt{Punturo2010a,Punturo2010b,sathyaprakash2012,Kalogera2019,GWIC3G}). 

A population of several hundred merging compact objects would be a Rosetta stone to address long-standing open questions in astrophysics \citep{stevenson2015,zevin2017,fishbach2017,fishbach2018,wysocki2018,bouffanais2019}: it can provide a key to understand  the evolution of massive binary stars \citep{bethe1998,belczynski2002,dominik2013, mennekens2014, spera2015,belczynski2016,marchant2016,mandel2016, eldridge2016, stevenson2017, chruslinska2018,kruckow2018,spera2019,Mapelli2019,eldridge2019,Giacobbo2018B,Giacobbo2019}, the formation channels of  compact binaries \citep{portegieszwart2000,  downing2010, rodriguez2015, rodriguez2018, askar2017, samsing2018a,samsing2018b, fragione2018, Gerosa2017,antonini2016, banerjee2010, mapelli2013,ziosi2014, mapelli2016, banerjee2017, banerjee2018, dicarlo2019, bouffanais2019}, and the role of primordial black holes as dark matter candidates \citep{carr1974,carr2016,Bird2016,Sasaki2016,Clesse2017}. Moreover, compact binary mergers can serve as standard sirens to estimate the Hubble constant \citep{Schutz1986,LIGO2017StandardSirens,Chen2018,Vitale2018,Fishbach2019,Soares-Santos2019,LIGO2019StandardSirens}.

Identifying the host galaxies of GW sources is crucial, in order to better understand the astrophysical implications of these events. For example, host galaxy identification is pivotal to measure the Hubble constant \citep{Fishbach2019,Gray2019} and gives important clues to reconstruct the formation channels of compact binaries (e.g. star clusters or field). So far, the host galaxy (NGC~4993) was identified only for the BNS merger GW170817  \citep{abbottmultimessenger,abbottGRB,goldstein2017,savchenko2017,margutti2017,coulter2017,soares-santos2017,chornock2017,cowperthwaite2017,nicholl2017,pian2017,alexander2017}.

Theoretical studies can give important contributions on the nature of host galaxies, eventually providing criteria to inform observational campaigns \citep[see e.g.,][]{OShaughnessy2010,Dvorkin2016,Lamberts2016,Schneider2017,Cao2018,Mapelli2018,Mapelli2018b,Artale2019,Marassi2019,Boco2019,Conselice2019,Mapelli2019}. Previous work suggests that the merger rate of compact objects per galaxy in the local Universe strongly correlates with the stellar mass and star-formation rate (SFR) of the host galaxy \citep[e.g.,][]{Mapelli2018,Cao2018,Artale2019}. Moreover, the typical stellar mass of the host galaxies of BNSs might be significantly different from that of merging BBHs and BHNSs, especially at low redshift \citep{Mapelli2018}. This difference comes from the interplay between  the distribution of delay times (i.e., the time between the formation of the stellar binary system and the merger event) and the dependence of BBH and BHNS merger efficiency on the metallicity of their progenitor stars \citep{Mapelli2018,Toffano2019}.

The host galaxies of merging compact objects can be also analysed in terms of early-type (ET) and late-type (LT) galaxies \citep{OShaughnessy2010,Artale2019}, which might help to disentangle the most likely places to detect GW events. Recently, \citet{Artale2019} proposed that ET galaxies host $\gtrsim{}60$ per cent, $\gtrsim{}64$ per cent, and $\gtrsim{}73$ per cent of all the BNSs, BHNSs, and BBHs merging in the local Universe.

In this work, we characterize the host galaxies of merging compact binaries across cosmic time (up to redshift $z\sim{}6$), by means of populations synthesis simulations 
\citep[run with \mobse{},][]{Giacobbo2018} 
combined with  galaxy catalogues from the hydrodynamical cosmological simulation \eagle\ \citep{Schaye15}. We show that the merger rate per galaxy strongly correlates with stellar mass, SFR and metallicity of the host galaxy, and we provide a simple fitting formula. These correlations mildly depend on redshift, tracing the evolution of the host galaxy population.

 This paper is organized as follows. \S~\ref{sec:method} describes the main properties of \eagle\ and {\sc mobse}, and the methodology implemented to seed the \eagle\ with a population of merging compact objects (BBHs, BNSs and BHNSs). We present our results in \S~\ref{sec:results}. Our main conclusions are discussed in  \S~\ref{sec:conclusions}.

\section{Method}\label{sec:method}

Following the same methodology originally introduced by \citet{Mapelli2017} \citep[see also][]{Mapelli2018,Mapelli2018b,Mapelli2019,Artale2019}, we
combine the results from the population synthesis code {\sc mobse} \citep{Giacobbo2018} with the galaxy catalogues from the \eagle\ simulation \citep{Schaye15}.
In this section we briefly describe the methodology; further details can be found in the aforementioned papers.

The population synthesis code {\sc mobse} \citep{Giacobbo2018} built from the {\sc bse} code \citep{Hurley2000,Hurley2002} includes new prescriptions for core collapse supernovae (SNe) based on \citet{Fryer2012}, metallicity dependent stellar winds \citep[see][]{Vink2001,Vink2005,Chen2015}, and pair-instability and pulsational pair-instability SNe \citep{Woosley2017,Spera2017}. Here we use the catalogue of merging compact objects from the run named as  CC15$\alpha{}$5 from \citet{Giacobbo2018B}.
This run was implemented also in \citet{Mapelli2018,Mapelli2019} and \citet{Artale2019}.
Run CC15$\alpha{}$5 is composed of 12 sub-sets of metallicities $Z = 0.0002$, 0.0004, 0.0008, 0.0012, 0.0016, 0.002, 0.004, 0.006, 0.008, 0.012, 0.016 and 0.02.
Each sub-set was run with $10^7$ stellar binaries, hence the total number of binary systems is $1.2\times10^{8}$. 
For each merging compact binary, the {\mobse} catalogue contains information about mass of the primary compact object, mass of the secondary compact object and delay time.

In this work we use  galaxy catalogues of two runs from the \eagle{} suite \citep[][]{Schaye15,Crain15}. The \eagle\ simulation, run with a modified version of the code {\sc gadget-3}, includes subgrid models to account for star formation, chemical enrichment, radiative cooling and heating,
UV/X-ray ionizing background, stellar evolution, AGB stars and SN feedback, and AGN feedback. 

We use the galaxy catalogue from the simulation {\sc RefL0100N1504} which represents a periodic box of $100\Mpc$
side (henceforth we refer to it as \eagle{}100) with initially $1504^3$ gas and dark matter particles. The initial mass of gas and dark matter particles is  $m_{\rm gas} = 1.81\times10^6 \Msun$ and $m_{\rm DM} = 9.70\times10^6 \Msun$, respectively. 
In order to test the robustness of our results, we also use the run named {\sc RecalL0025N0752run}, which represents a periodic box of $25\Mpc$ side (we refer to it as \eagle{}25). \eagle{}25 was run with the highest resolution in the \eagle\ suite; it initially contains $752^3$ gas and dark matter particles with $m_{\rm gas} = 2.26\times10^5 \Msun$ and $m_{\rm DM} = 1.21\times10^6 \Msun$. 
The two simulated boxes are complementary, since \eagle{}25 has the highest resolution in the \eagle\ suite \citep{McAlpine2016}, while \eagle{}100 contains massive galaxies with stellar masses up to $1.9\times10^{12}\Msun$. Hence, by comparing the results of \eagle{}25 and \eagle{}100 we can estimate the robustness of our findings with respect to resolution and box-size related issues. 

In the main text we discuss the results of \eagle{}100 only. We present the results of \eagle{}25 and compare them with \eagle{}100 in Appendix~\ref{sec:appConv}. We find very good agreement between the results we obtain with \eagle{}100 and those with \eagle{}25.

Information on the simulated galaxies is available on the {\sc sql} database\footnote{{\url http://icc.dur.ac.uk/Eagle/}, {\url http://virgo.dur.ac.uk/data.php}.}. The \eagle{} suite adopts the $\Lambda$CDM cosmology with parameters inferred from \cite{Planck13} ($\Omega_{\rm m} = 0.2588$, $\Omega_\Lambda = 0.693$,
$\Omega_{\rm b} = 0.0482$, and $H_0 = 100\,{} h$ km s$^{-1}$ Mpc$^{-1}$ with $h = 0.6777$). The simulation was run from $z = 127$ up to $z \sim 0$.
From the database, we can extract the properties of the host galaxies (e.g. stellar mass $M_\ast{}$, SFR and metallicity of star-forming gas).

With a Monte Carlo algorithm, we assign a number of compact binaries (from {\sc mobse} simulations) to each \eagle{} stellar particle, based on its initial mass and metallicity. The formation time of compact-binary progenitors assigned to a given \eagle{} particle is assumed to be the same as the formation time of the \eagle{} particle. Each compact binary merges in the same stellar particle to which it was assigned (we do not allow for ejections of compact binaries) after  its individual delay time (estimated from {\sc mobse}). 
Following this procedure, we assign to each \eagle{} galaxy  a population of merging compact objects across cosmic time.

In this work, we study the host galaxies of BBHs, BHNSs, and BNSs at four representative redshift intervals: $z=0-0.1$ \citep[studied also in][]{Artale2019},
$z=0.93-1.13$, $z=1.87-2.12$ , and $z=5.73-6.51$.  Hereafter, we will refer to these redshift intervals as $z=0.1$, $z=1$, $z=2$ and $z=6$ for brevity. The first interval ($z=0.1$) is our local Universe sample, while $z=1$ is representative of the advanced LIGO and Virgo horizon for BBHs at design sensitivity. Redshift $z=2$ corresponds to the peak of star formation, while redshift $z=6$ is approximately the end of the reionization epoch (we do not consider higher redshifts because the \eagle{} does not include prescriptions for population~III stars). 

We select only galaxies with stellar mass above $10^7~\Msun$ and ${\rm SFR} > 0\,{}\Msun{}$ yr$^{-1}$. Hence, the catalogue from \eagle{}100 has in total 77959, 91294, 116074, and 50544 galaxies at $z = 0.1, 1, 2,$ and 6, respectively. In the \eagle{}25, we get 3070, 1827, 1806, and 412 galaxies at $z = 0.1, 1, 2$, and 6, respectively.

\section{Results}\label{sec:results}
\begin{table*} 
  \caption{Fits of the BNS merger rate per galaxy, from \eagle{}100 
    at $z = 0.1, 1, 2$, and 6  (see Sec.~\ref{sec:results_fits} for details).}\label{table:Fits100MpcDNS}
\begin{tabular}{cc|c|c|c|c|l}
\cline{3-6}
& & \multicolumn{4}{ c| }{BNSs} \\ \cline{3-6}
& & $z = 0.1$ & $z=1$ & $z=2$ & $z=6$ \\ \cline{1-6}
\multicolumn{1}{ |c  }{\multirow{2}{*}{Fit~1D} } &
\multicolumn{1}{ |c| }{$\alpha_{1}$} & $1.038   \pm  0.001 $  & $1.109   \pm  0.001 $   & $1.050   \pm  0.001 $  & $1.027   \pm  0.003 $   &     \\ \cline{2-6}
\multicolumn{1}{ |c  }{}                        &
\multicolumn{1}{ |c| }{ $\alpha_{2}$} & $-6.090   \pm 0.010  $  & $-6.214   \pm  0.006 $   & $-5.533   \pm  0.006 $   & $-5.029   \pm  0.021 $   &     \\ \cline{1-6}
\multicolumn{1}{ |c  }{\multirow{2}{*}{Fit~2D} } &
\multicolumn{1}{ |c| }{$\beta_{1}$} & $0.800   \pm  0.002 $  & $0.964   \pm  0.001 $  & $0.977   \pm 0.001  $   & $1.113   \pm  0.003 $   &  \\ \cline{2-6}
\multicolumn{1}{ |c  }{}                        &
\multicolumn{1}{ |c| }{$\beta_{2}$} & $0.323   \pm  0.002 $  & $0.155   \pm  0.001 $  & $0.068   \pm  0.001 $   & $-0.070   \pm  0.002 $   &  \\ \cline{2-6}
\multicolumn{1}{ |c  }{}                        &
\multicolumn{1}{ |c| }{$\beta_{3}$} & $-3.555   \pm  0.018 $  & $-4.819   \pm  0.013 $   & $-4.874   \pm  0.011 $   & $-5.764   \pm  0.026 $   &  \\ \cline{1-6}
\multicolumn{1}{ |c  }{\multirow{2}{*}{Fit~3D} } &
\multicolumn{1}{ |c| }{$\gamma_{1}$} & $0.701   \pm  0.002 $  & $0.896   \pm  0.002 $   & $1.018   \pm   0.002$   & $1.137   \pm  0.004 $   &  \\ \cline{2-6}
\multicolumn{1}{ |c  }{}                        &
\multicolumn{1}{ |c| }{$\gamma_{2}$} & $0.356   \pm  0.002 $  & $0.184   \pm  0.001 $  & $0.048   \pm   0.001$   & $-0.082   \pm  0.002 $   & \\ \cline{2-6}
\multicolumn{1}{ |c  }{}                        &
\multicolumn{1}{ |c| }{$\gamma_{3}$} & $0.411   \pm  0.005 $  &  $0.222   \pm 0.003  $ & $-0.103   \pm  0.002 $   & $-0.053   \pm  0.004 $   & \\ \cline{2-6}
\multicolumn{1}{ |c  }{}                        &
\multicolumn{1}{ |c| }{$\gamma_{4}$} & $-1.968   \pm  0.026 $  & $-3.795   \pm  0.019 $  & $-5.451   \pm   0.017$  & $-6.104   \pm  0.037 $   & \\ \cline{1-6}
\end{tabular}
\end{table*}
\begin{table*}
\caption{Same as Table~\ref{table:Fits100MpcDNS} but for BHNSs.}\label{table:Fits100MpcBHNS}
\begin{tabular}{cc|c|c|c|c|l}
\cline{3-6}
& & \multicolumn{4}{ c| }{BHNSs} \\ \cline{3-6}
& & $z = 0.1$ & $z=1$ & $z=2$ & $z=6$ \\ \cline{1-6}
\multicolumn{1}{ |c  }{\multirow{2}{*}{Fit~1D} } &
\multicolumn{1}{ |c| }{$\alpha_{1}$} & $0.824   \pm  0.001 $  & $0.873   \pm   0.001$   & $0.913   \pm  0.001 $   & $0.965   \pm  0.002 $  &     \\ \cline{2-6}
\multicolumn{1}{ |c  }{}                        &
\multicolumn{1}{ |c| }{ $\alpha_{2}$} & $-4.731   \pm  0.008 $  & $-4.478   \pm  0.008 $   & $-4.401   \pm  0.007 $   & $-4.315   \pm  0.018 $   &     \\ \cline{1-6}
\multicolumn{1}{ |c  }{\multirow{2}{*}{Fit~2D} } &
\multicolumn{1}{ |c| }{$\beta_{1}$} & $0.711   \pm  0.002 $  & $0.813   \pm 0.002  $   & $0.871   \pm  0.002 $   & $0.985   \pm  0.003 $   &  \\ \cline{2-6}
\multicolumn{1}{ |c  }{}                        &
\multicolumn{1}{ |c| }{$\beta_{2}$} & $0.150   \pm  0.002 $   & $0.064   \pm  0.002 $  & $0.039   \pm  0.001 $   & $-0.017   \pm 0.001  $   &  \\ \cline{2-6}
\multicolumn{1}{ |c  }{}                        &
\multicolumn{1}{ |c| }{$\beta_{3}$} & $-3.536   \pm   0.016$   & $-3.900   \pm  0.018 $   & $-4.019   \pm  0.014 $   & $-4.490   \pm  0.024 $   &  \\ \cline{1-6}
\multicolumn{1}{ |c  }{\multirow{2}{*}{Fit~3D} } &
\multicolumn{1}{ |c| }{$\gamma_{1}$} & $0.833   \pm  0.002 $   & $1.074   \pm  0.002 $   & $1.084   \pm  0.002 $   & $0.978   \pm 0.003  $   &  \\ \cline{2-6}
\multicolumn{1}{ |c  }{}                        &
\multicolumn{1}{ |c| }{$\gamma_{2}$} & $0.101   \pm   0.002$   & $-0.058   \pm  0.002 $   & $-0.068   \pm  0.001 $   & $-0.013   \pm 0.002  $   & \\ \cline{2-6}
\multicolumn{1}{ |c  }{}                        &
\multicolumn{1}{ |c| }{$\gamma_{3}$} & $-0.461   \pm  0.004 $  & $-0.788   \pm   0.004$  & $-0.535   \pm   0.003$   & $0.016   \pm  0.004 $    & \\ \cline{2-6}
\multicolumn{1}{ |c  }{}                        &
\multicolumn{1}{ |c| }{$\gamma_{4}$} & $-5.434   \pm   0.022$  & $-7.733   \pm  0.023 $  & $-7.055   \pm  0.018 $   & $-4.386   \pm  0.034 $   & \\ \cline{1-6}
\end{tabular}
\end{table*}
\begin{table*}
\caption{Same as Table~\ref{table:Fits100MpcDNS} but for BBHs.}\label{table:Fits100MpcDBH}
\begin{tabular}{cc|c|c|c|c|l}
\cline{3-6}
& & \multicolumn{4}{ c| }{BBHs} \\ \cline{3-6}
& & $z = 0.1$ & $z=1$ & $z=2$ & $z=6$ \\ \cline{1-6}
\multicolumn{1}{ |c  }{\multirow{2}{*}{Fit~1D} } &
\multicolumn{1}{ |c| }{$\alpha_{1}$} & $0.807   \pm  0.001 $  & $0.813   \pm  0.001 $   & $0.831   \pm   0.001$  & $0.933   \pm  0.004 $   &     \\ \cline{2-6}
\multicolumn{1}{ |c  }{}                        &
\multicolumn{1}{ |c| }{ $\alpha_{2}$} & $-4.310   \pm  0.006 $  & $-3.845   \pm  0.008 $   & $-3.600   \pm  0.008 $   & $-4.190   \pm  0.026 $   &     \\ \cline{1-6}
\multicolumn{1}{ |c  }{\multirow{2}{*}{Fit~2D} } &
\multicolumn{1}{ |c| }{$\beta_{1}$} & $0.812   \pm   0.001$  & $0.840   \pm   0.002$   & $0.858   \pm  0.002 $   & $1.053   \pm  0.004 $   &  \\ \cline{2-6}
\multicolumn{1}{ |c  }{}                        &
\multicolumn{1}{ |c| }{$\beta_{2}$} & $-0.006   \pm  0.001 $  & $-0.029   \pm  0.002 $  & $-0.026   \pm  0.001 $   & $-0.098   \pm   0.002$  &  \\ \cline{2-6}
\multicolumn{1}{ |c  }{}                        &
\multicolumn{1}{ |c| }{$\beta_{3}$} & $-4.358   \pm  0.013 $   & $-4.109   \pm 0.018  $   & $-3.850   \pm   0.015$   & $-5.213   \pm  0.034 $   &  \\ \cline{1-6}
\multicolumn{1}{ |c  }{\multirow{2}{*}{Fit~3D} } &
\multicolumn{1}{ |c| }{$\gamma_{1}$} & $0.921   \pm   0.001$   & $1.134   \pm  0.002 $   & $1.135   \pm   0.002$   & $1.131   \pm  0.005 $   &  \\ \cline{2-6}
\multicolumn{1}{ |c  }{}                        &
\multicolumn{1}{ |c| }{$\gamma_{2}$} & $-0.051   \pm  0.001 $   & $-0.172   \pm  0.002 $   & $-0.167   \pm  0.001 $   & $-0.137   \pm  0.002 $   & \\ \cline{2-6}
\multicolumn{1}{ |c  }{}                        &
\multicolumn{1}{ |c| }{$\gamma_{3}$} & $-0.404   \pm  0.003 $   & $-0.839   \pm  0.004 $  & $-0.681   \pm  0.003 $   & $-0.171   \pm   0.005$   & \\ \cline{2-6}
\multicolumn{1}{ |c  }{}                        &
\multicolumn{1}{ |c| }{$\gamma_{4}$} & $-6.049   \pm  0.018 $  & $-8.338   \pm  0.024 $   & $-7.758   \pm  0.020 $   & $-6.321   \pm 0.047  $   & \\ \cline{1-6}
\end{tabular}
\end{table*}
\subsection{Merger rate per galaxy}
\label{sec:results_fits}

For each redshift interval, we calculate the number of BNSs, BHNSs and BBHs merging in each simulated galaxy. This number, divided by the considered time-span (1.84, 0.76, 0.41, and 0.14~Gyr for $z=0.1$, 1, 2, and 6, respectively), gives the merger rate  per galaxy at a given redshift ($n_{\rm GW}$).
\cite{Artale2019} already computed this quantity for the local Universe, using the galaxy catalog from \eagle{}25. Here, we extend their analysis to higher redshift ($z=1,$ 2 and 6) and using a larger simulated box.

We then performed a series of fits of increasing complexity and dimensionality that we refer to as Fit~1D, Fit~2D and Fit~3D, with the following prescriptions:\\
Fit~1D: we fit the merger rate per galaxy as a function of stellar mass ($M_{\ast{}}$) only, as
  \begin{equation}
    \log(n_{\rm GW}/{\rm Gyr}) = \alpha_1 \log(M_{\ast{}}/ \Msun) + \alpha_2.
    \end{equation}
Fit~2D: we fit the merger rate per galaxy as a function of both $M_{\ast{}}$ and SFR, as
  \begin{eqnarray}
    \log(n_{\rm GW}/{\rm Gyr}) = \beta_1 \log(M_{\ast{}}/ \Msun) \nonumber\\
    + \beta_2 \log({\rm SFR} / \Msun \yr^{-1}) + \beta_3.
    \end{eqnarray}
Fit~3D:  we fit the merger rate per galaxy as a function of $M_{\ast{}}$, SFR and metallicity $Z$, as
  \begin{eqnarray}
    \log(n_{\rm GW}/{\rm Gyr}) = \gamma_1 \log(M_{\ast{}}/ \Msun) \nonumber\\
    + \gamma_2 \log({\rm SFR} / \Msun \yr^{-1}) + \gamma_3 \log(Z) + \gamma_4.
    \end{eqnarray}

We adopted a standard linear regression approach using least-squares estimation to derive the values of the coefficients and their standard deviations. The results obtained are presented  in Table~~\ref{table:Fits100MpcDNS}, \ref{table:Fits100MpcBHNS} and ~\ref{table:Fits100MpcDBH} for BNSs, BHNSs and BBHs respectively. These fits can help to identify the most likely host galaxies of merging compact objects according to their SFR, stellar mass and metallicity.

\begin{figure*}
\includegraphics[width=2\columnwidth]{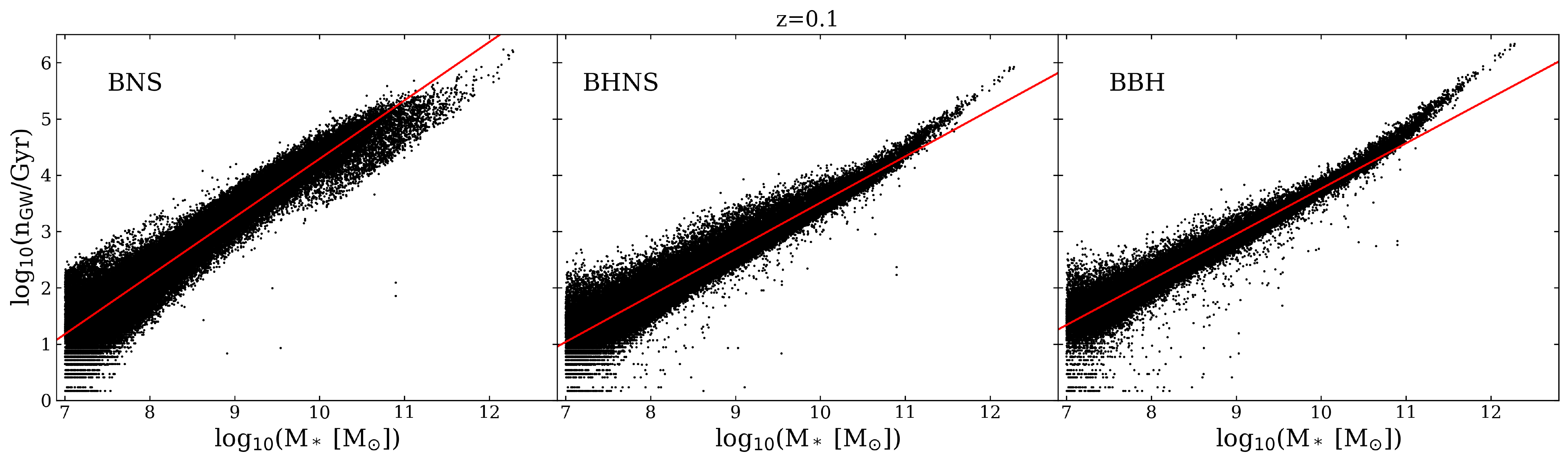}
\includegraphics[width=2\columnwidth]{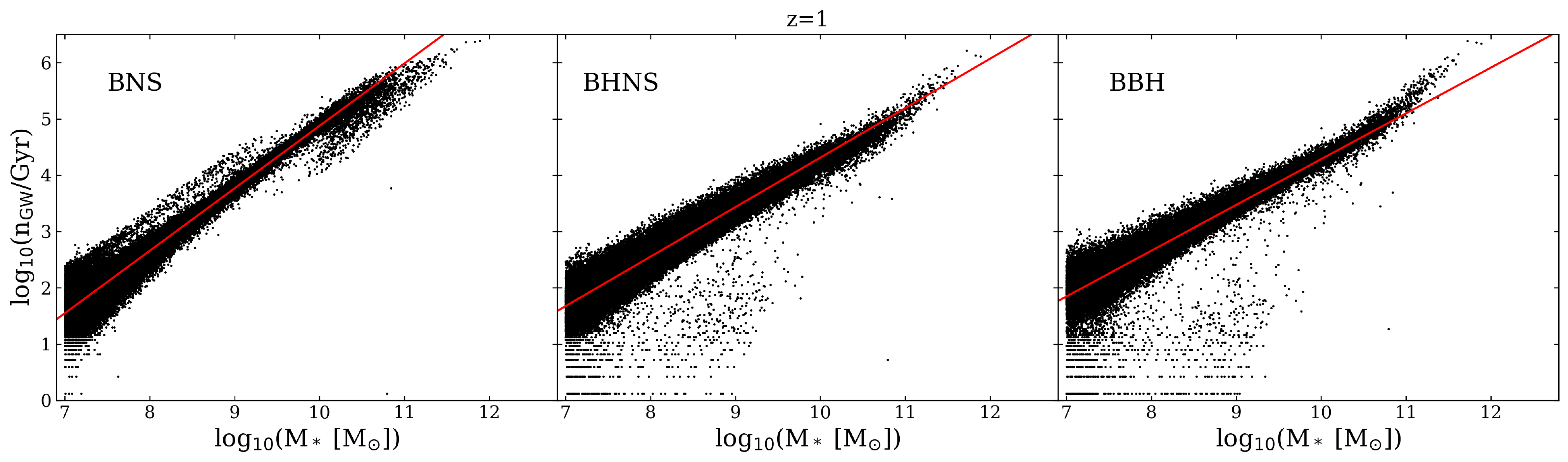}
\includegraphics[width=2\columnwidth]{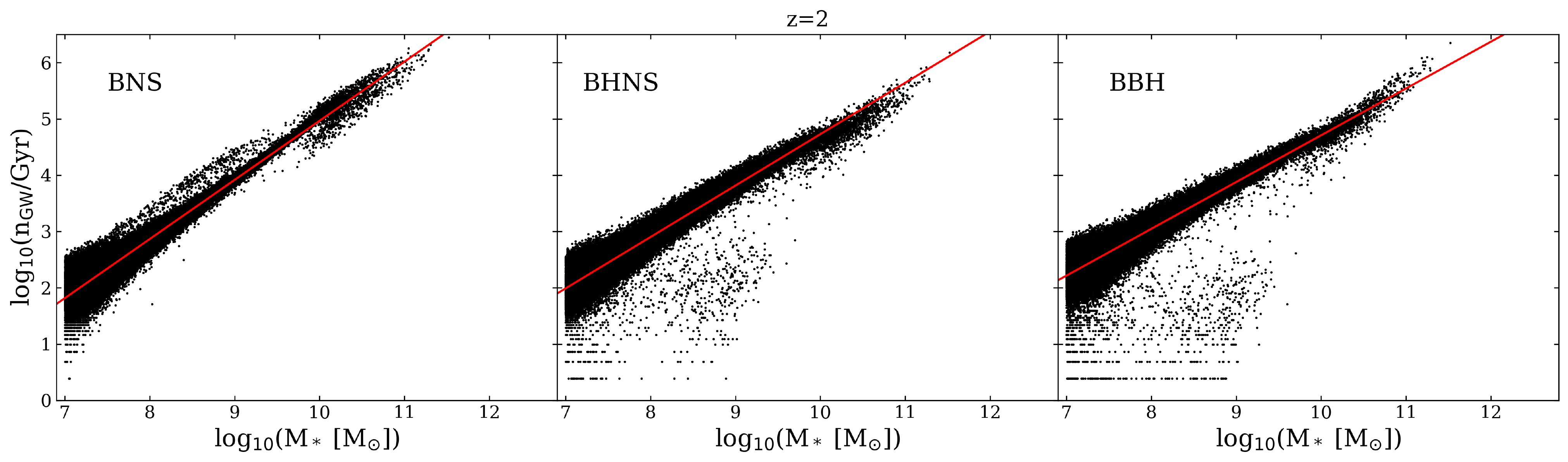}
\includegraphics[width=2\columnwidth]{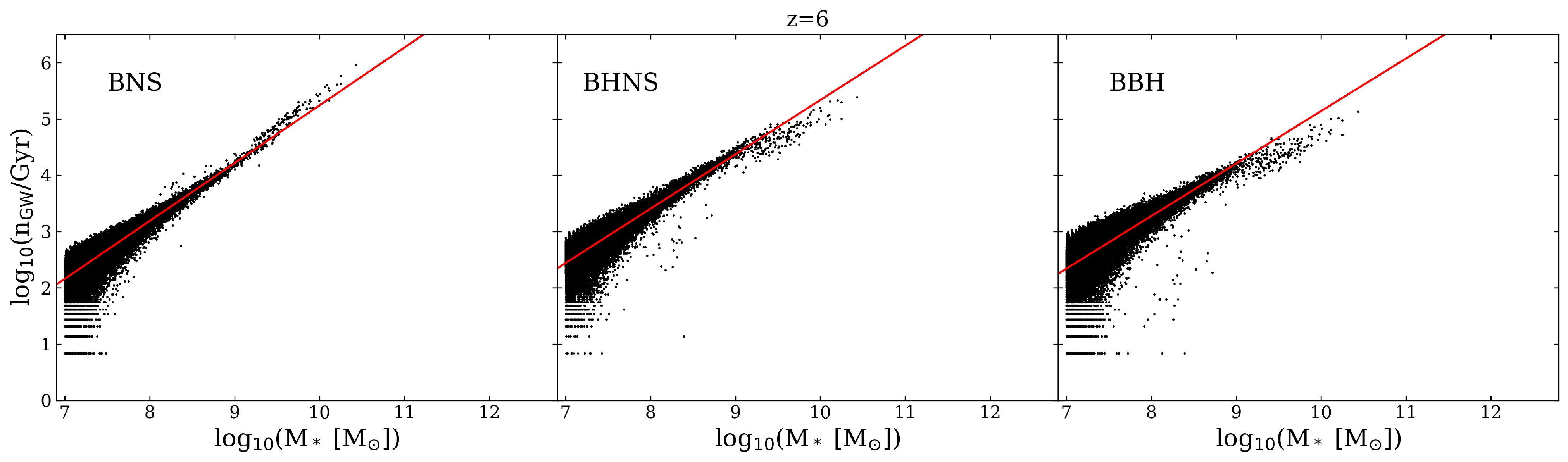}
\caption{Merger rate per galaxy (n$_{\rm GW}$) as a function of stellar mass  $M_\ast{}$ for the host galaxies of merging BNSs (left-hand panels), BHNSs (middle panels), and BBHs (right-hand panels), using \eagle{}100. From top to bottom row: $z=0.1$, $z=1$, $z=2$ and $z=6$. Red lines: Fit~1D (see Tables~\ref{table:Fits100MpcDNS}, \ref{table:Fits100MpcBHNS} and \ref{table:Fits100MpcDBH}).}
\label{fig:nGW_Ms_100Mpc}
\end{figure*}

Figure~\ref{fig:nGW_Ms_100Mpc} shows the evolution of the merger rate per galaxy as a function of the stellar mass at $z=0.1, 1, 2,$ and 6, for BNS (left column), BHNS (middle column) and BBHs (right column). In all cases, we have a strong visual correlation between $n_{\rm GW}$ and the stellar mass of the host galaxy \citep[in agreement with the correlation found by][at $z\lesssim{}0.1$]{Artale2019}, that is supported by Fit~1D represented by the red line on Figure ~\ref{fig:nGW_Ms_100Mpc}. This correlation holds with relatively minor changes for all considered redshifts and is steeper for merging BNSs than for BBHs and BHNSs. 

In addition, we find that adding the SFR and metallicity in Fit~2D and Fit~3D helped improve the quality of our fit. Quantitatively, we performed a simple model selection by computing the Bayesian Information Criterion (BIC) for all models \citep{schwarz1978}, and found that $\text{BIC}_{\text{1D}}>\text{BIC}_{\text{2D}}>\text{BIC}_{\text{3D}}$, indicating that the fit with more parameters is always preferred. It is worth mentioning that this behaviour is expected as the number of points used to do these fits is quite high and the dimensionality of our fits is low.

For the host galaxies of merging BBHs at $z=0.1$, we find that the relation between the merger rate per galaxy and the stellar mass has a slight change in the slope at $\log(M_\ast{}/M_\odot) \sim 10.5$ (see Fig.~\ref{fig:nGW_Ms_100Mpc}). This is also seen but less significant at $z=1$, and 2.
The trend is not present for the BNS hosts, and it is subtle for BHNS hosts. 

The slope change for the host galaxies of merging BBHs is explained  by the interplay between the mass-metallicity relation (MZR) of the galaxies and the strong dependence that BBH progenitors have on metallicity. 
Observational and numerical results have shown that the MZR is steep for galaxies with $M_\ast{} \sim 10^{8.5} - 10^{10.5} \Msun$, with a turnover for $M_\ast{} > 10^{10.5} \Msun$ \citep[see e.g.,][]{Tremonti2004,Creasey2015}.
This is understood to be the result of the role that stellar feedback and AGN feedback have on the chemical evolution of galaxies. In particular the subgrid model from \eagle{} suite (in both, \eagle{}100 and \eagle{}25) accounts for these feedbacks and shows the turnover in the MZR \citep{Segers2016}.   
We quantify the change in the slope and provide more details in Appendix~\ref{sec:galprop_mergRates}.

\subsection{Merger probability per a given host's stellar mass and SFR}\label{sec:JointProb}

\begin{figure*}
\includegraphics[width=\columnwidth]{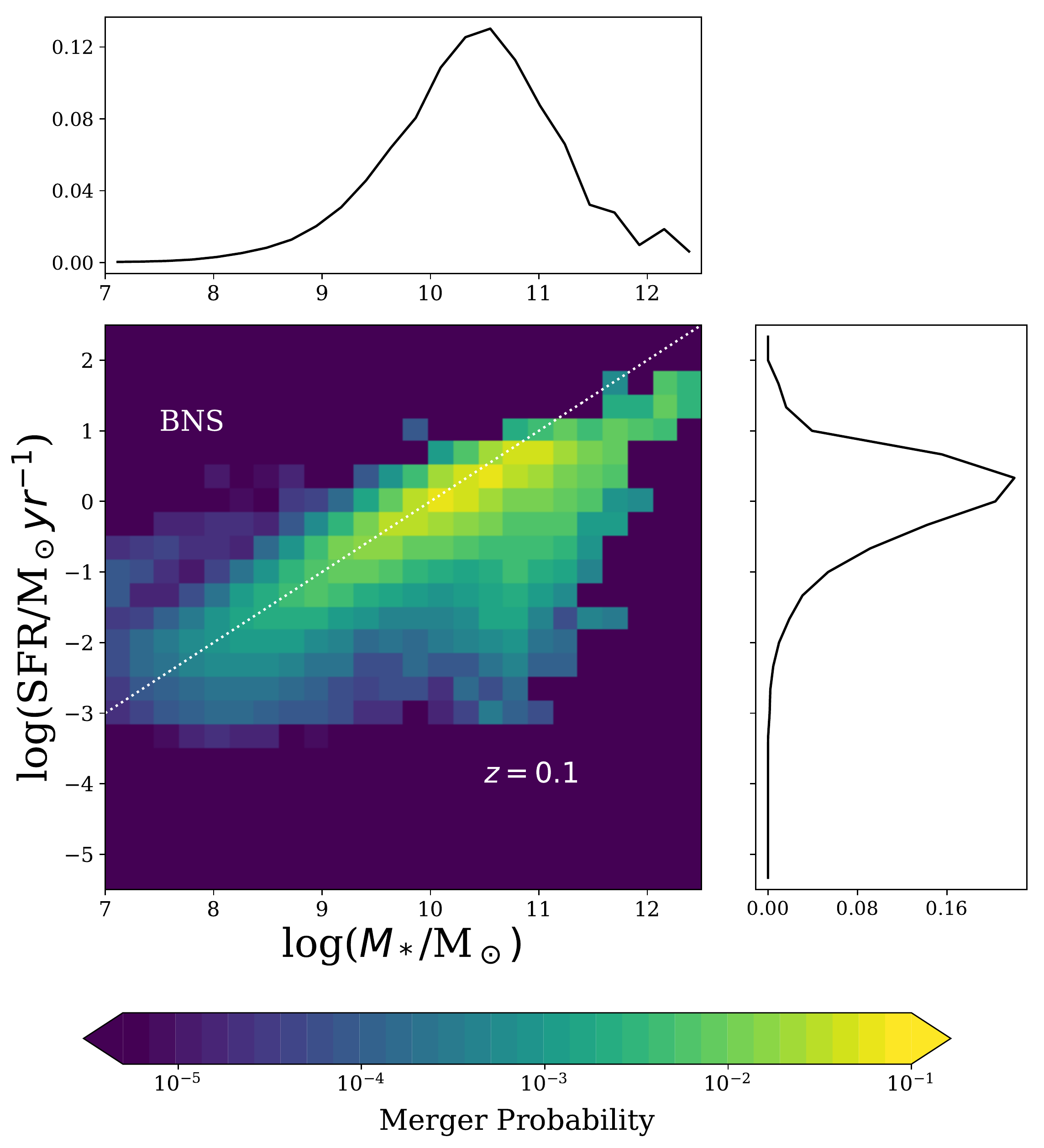}
\includegraphics[width=\columnwidth]{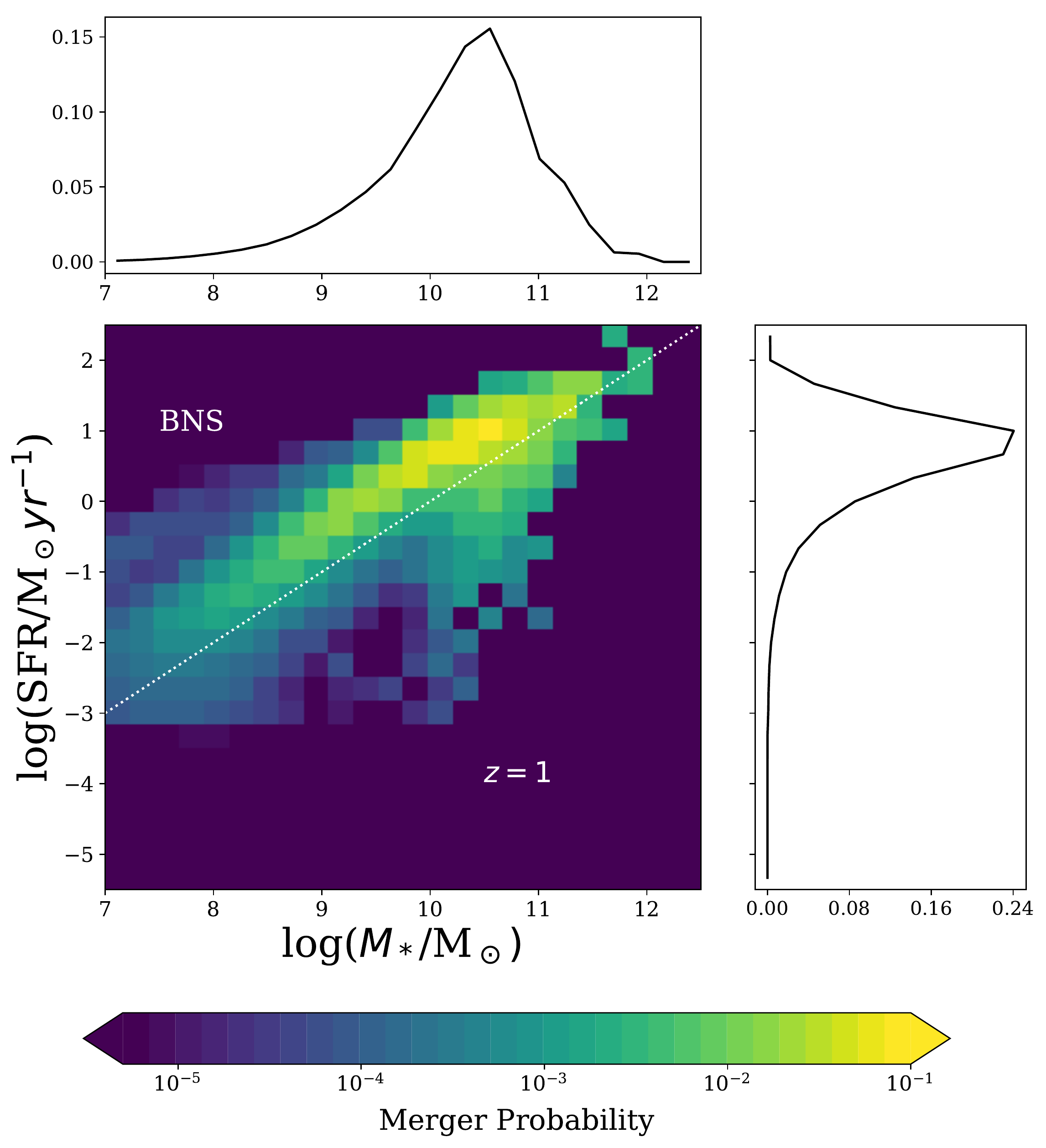}
\includegraphics[width=\columnwidth]{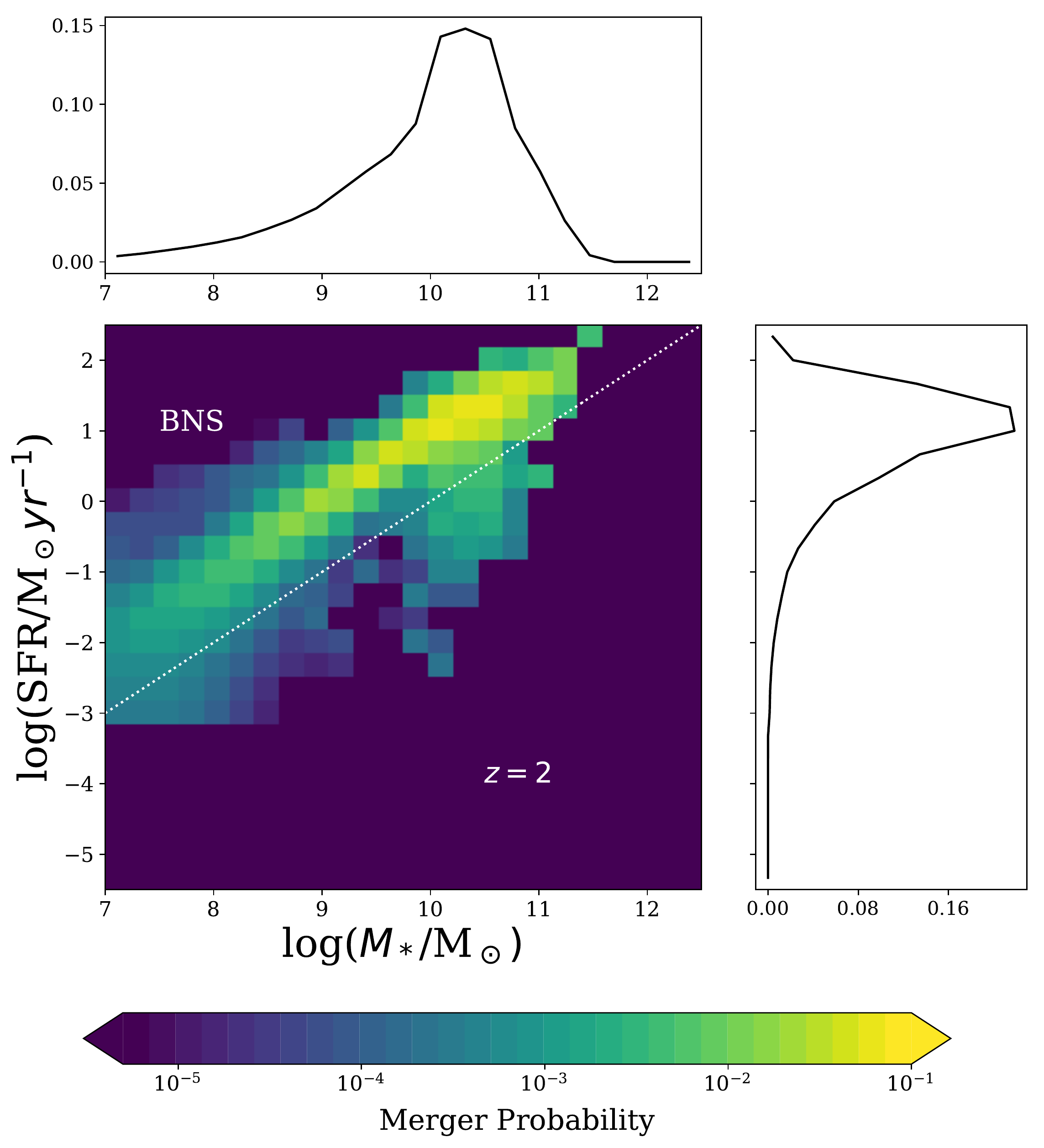}
\includegraphics[width=\columnwidth]{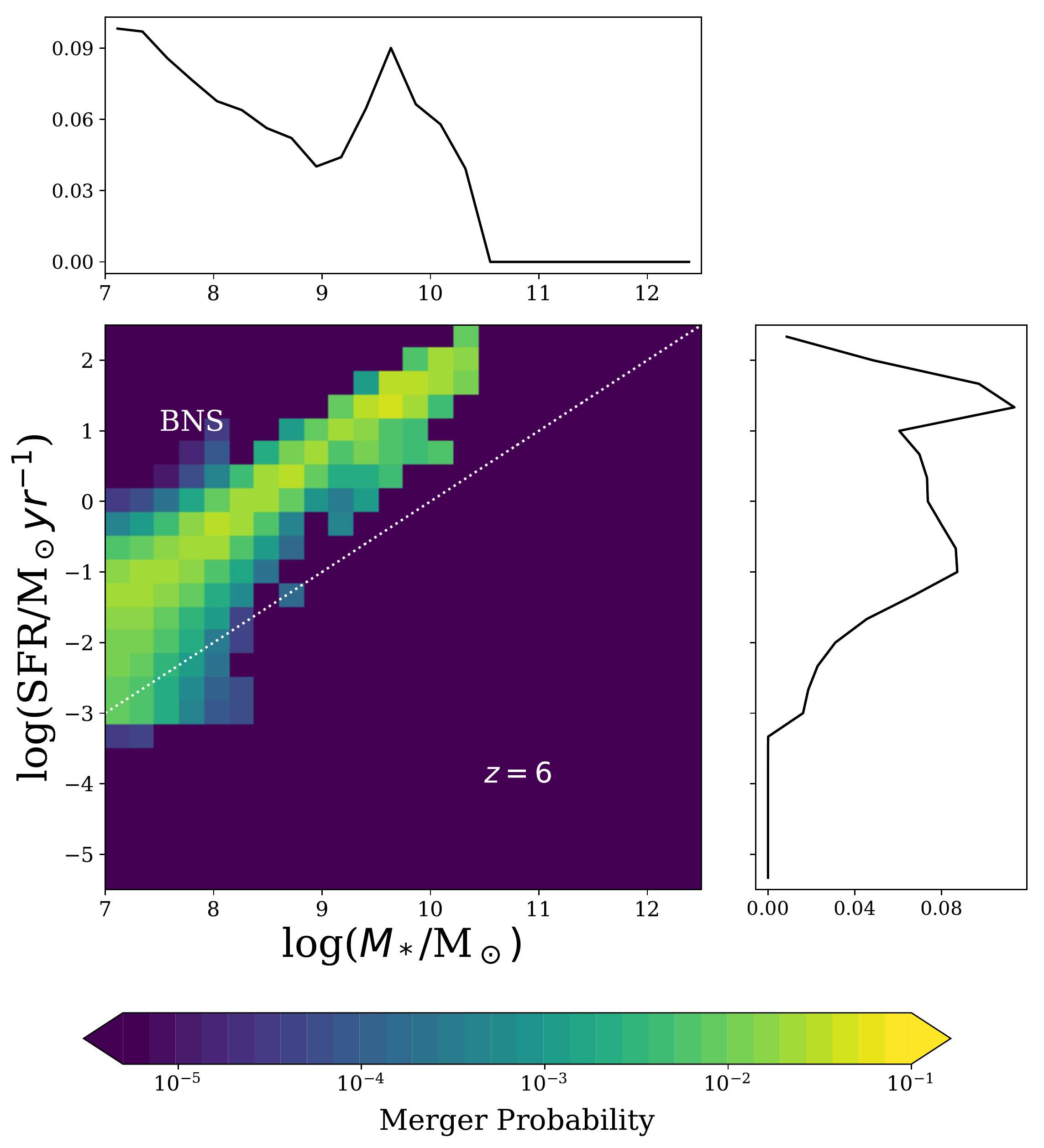}
\caption{Merger probability per a given SFR and stellar mass ($M_\ast$) bin for BNSs, estimated from \eagle{}100. The merger probability is calculated as described in equation~\ref{eq:prob}. From top to bottom and from left to right:  $z = 0.1$, $z= 1, 2$ and 6.
The white dotted line represents a specific SFR (sSFR) $ = 10^{-10}$~yr$^{-1}$. The marginal plots represent the distributions of SFR and $M_\ast{}$.} 
\label{fig:DNS_prob_100Mpc}
\end{figure*}
\begin{figure*}
\includegraphics[width=\columnwidth]{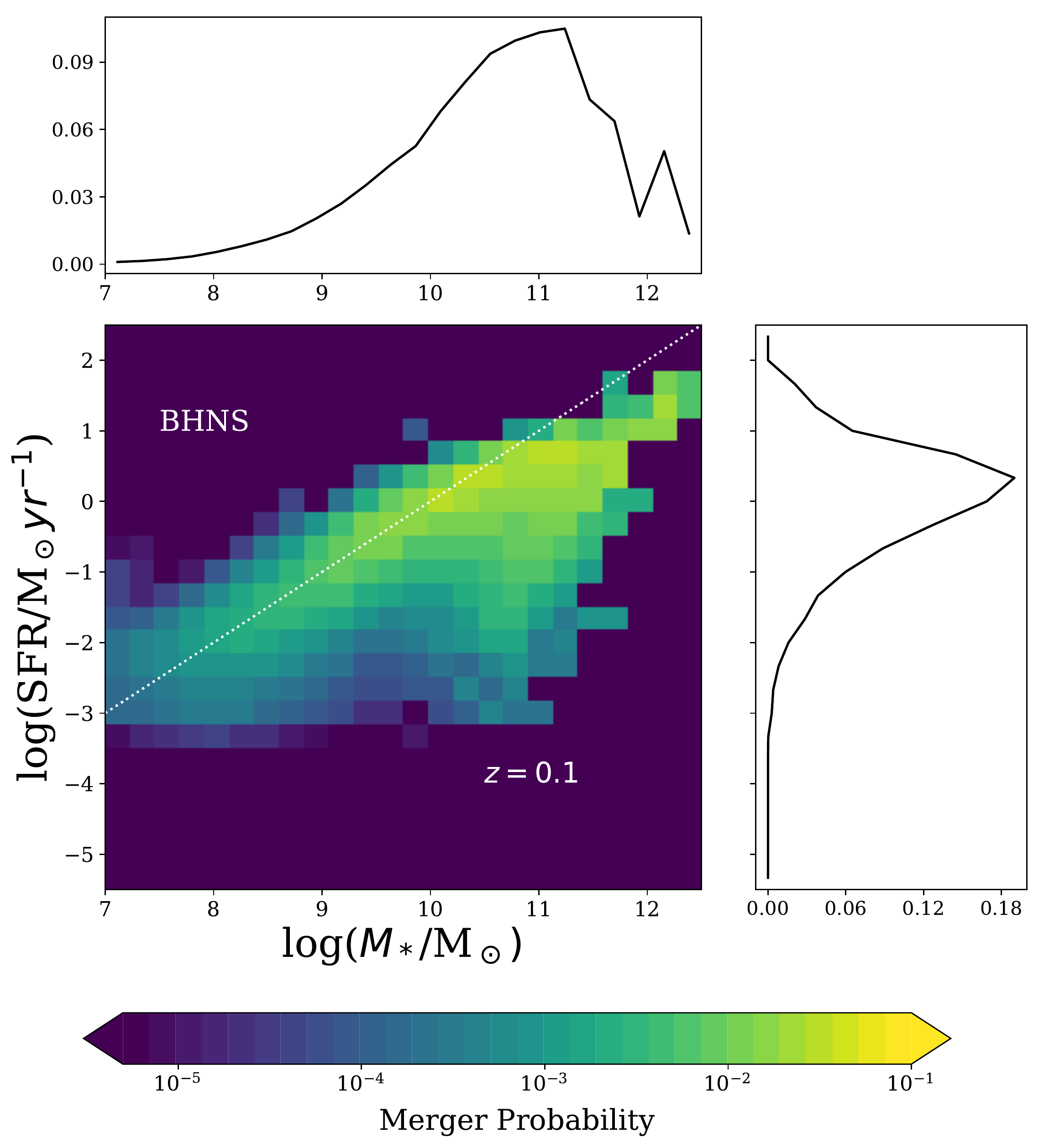}
\includegraphics[width=\columnwidth]{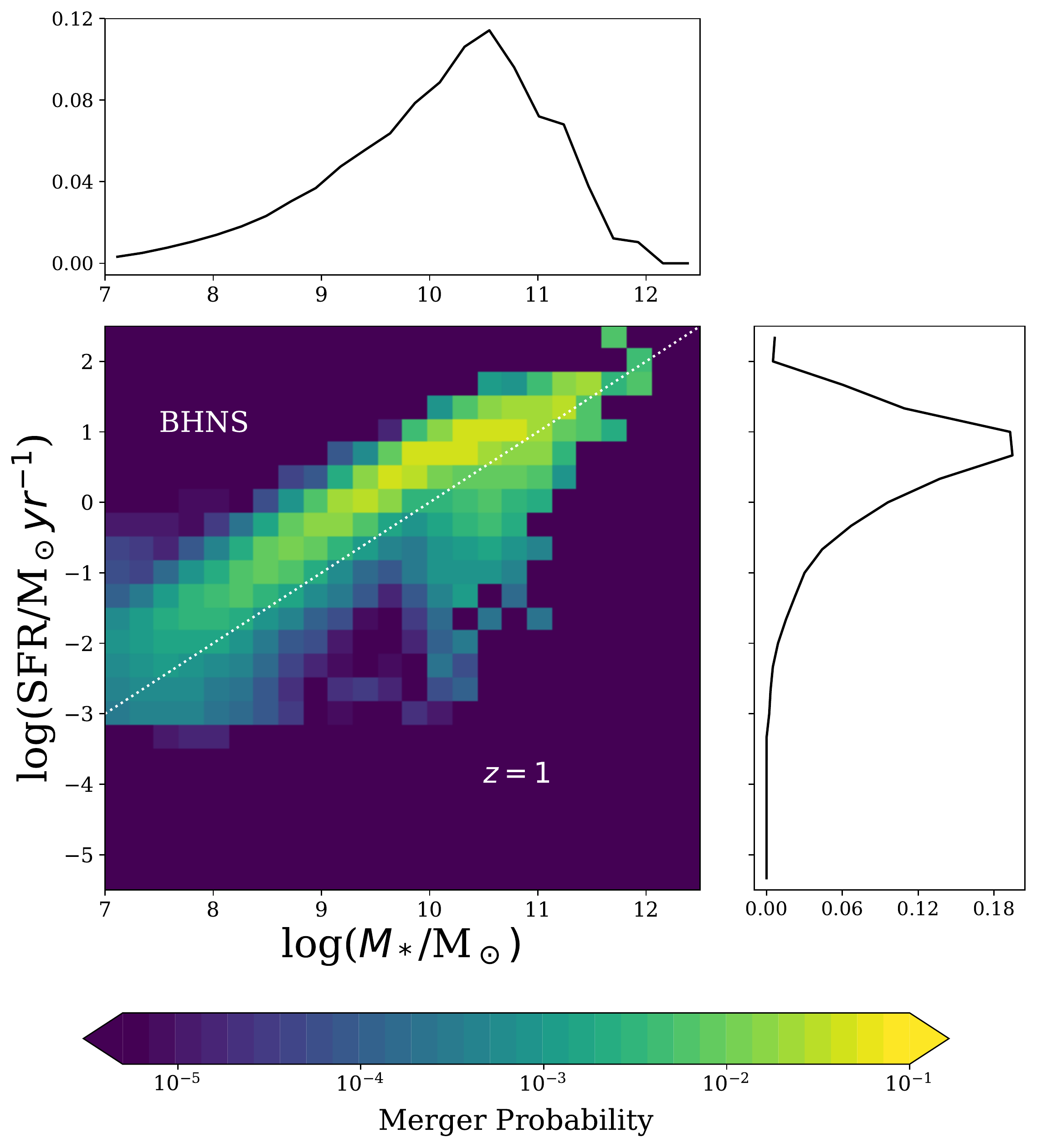}
\includegraphics[width=\columnwidth]{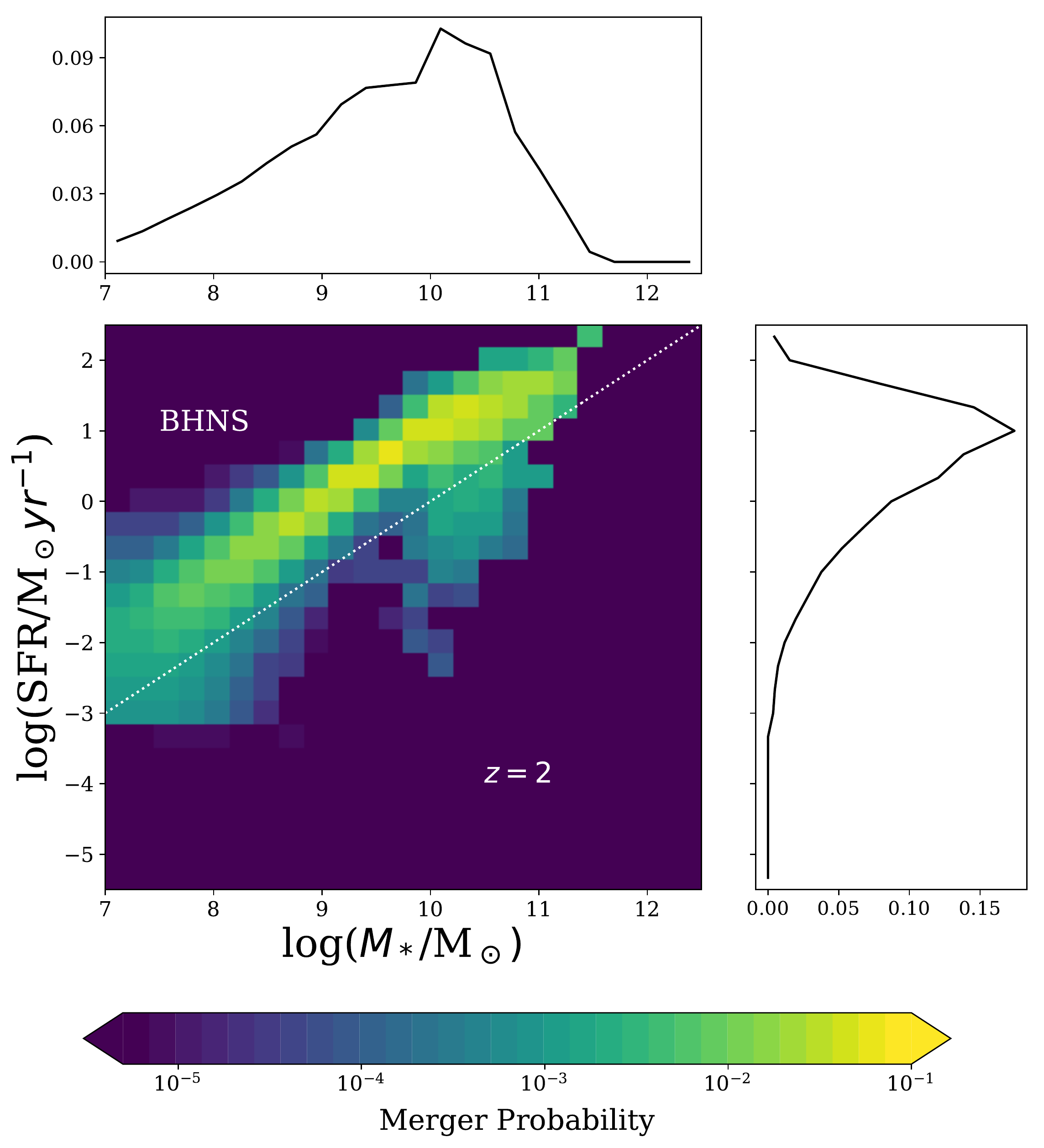}
\includegraphics[width=\columnwidth]{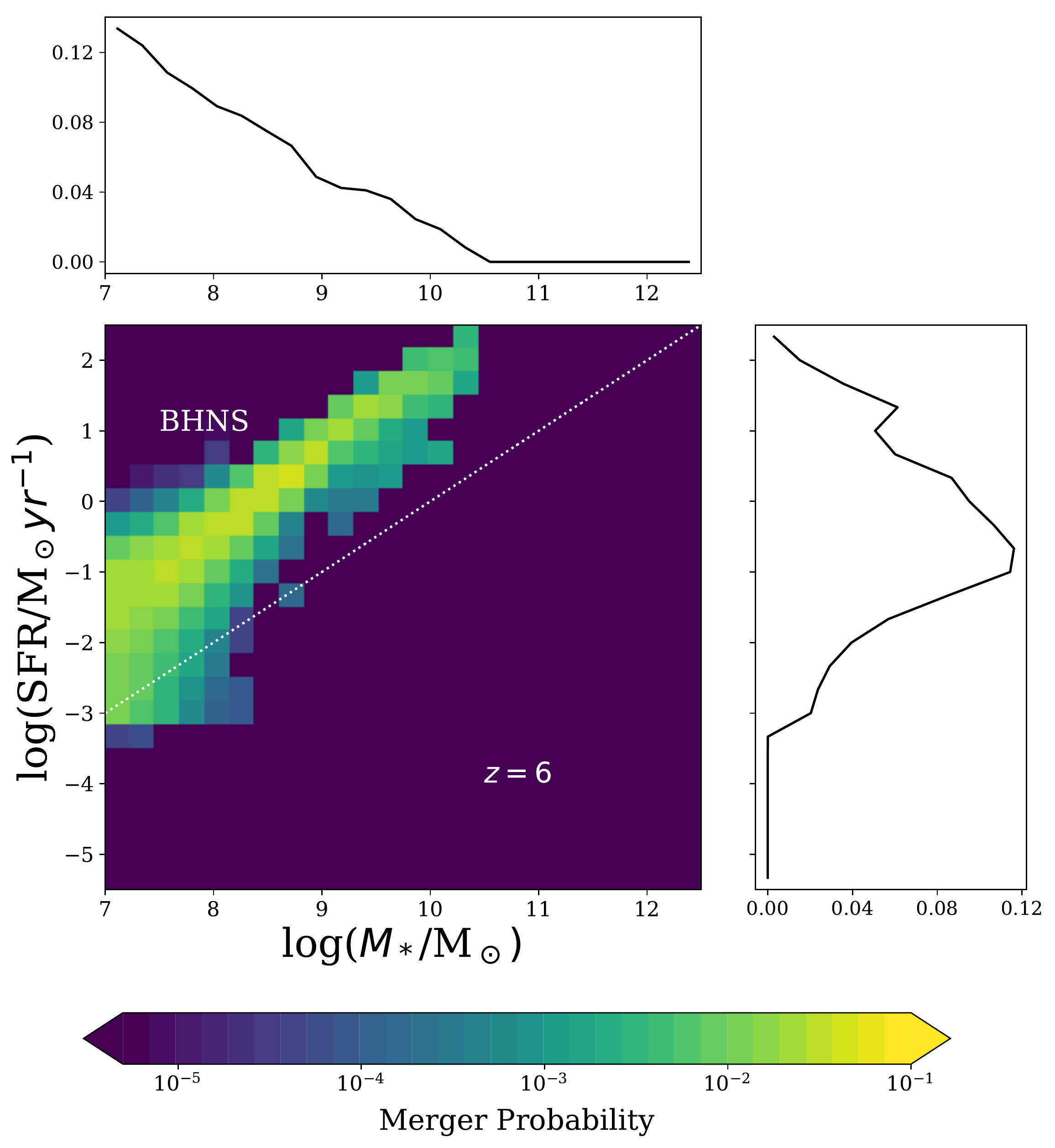}
\caption{Same as Figure~\ref{fig:DNS_prob_100Mpc}, but for BHNSs.}
\label{fig:BHNS_prob_100Mpc}
\end{figure*}
\begin{figure*}
\includegraphics[width=\columnwidth]{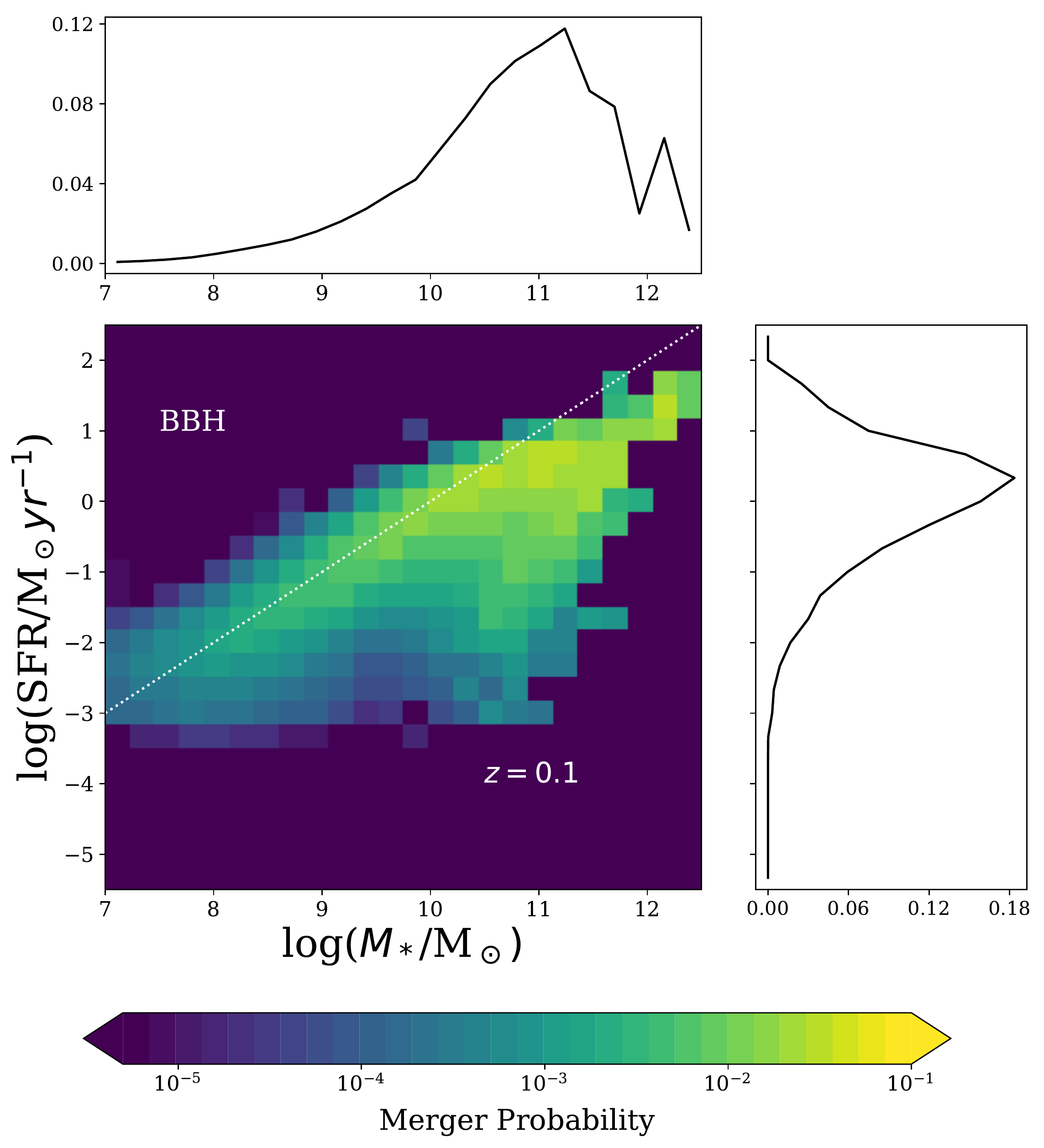}
\includegraphics[width=\columnwidth]{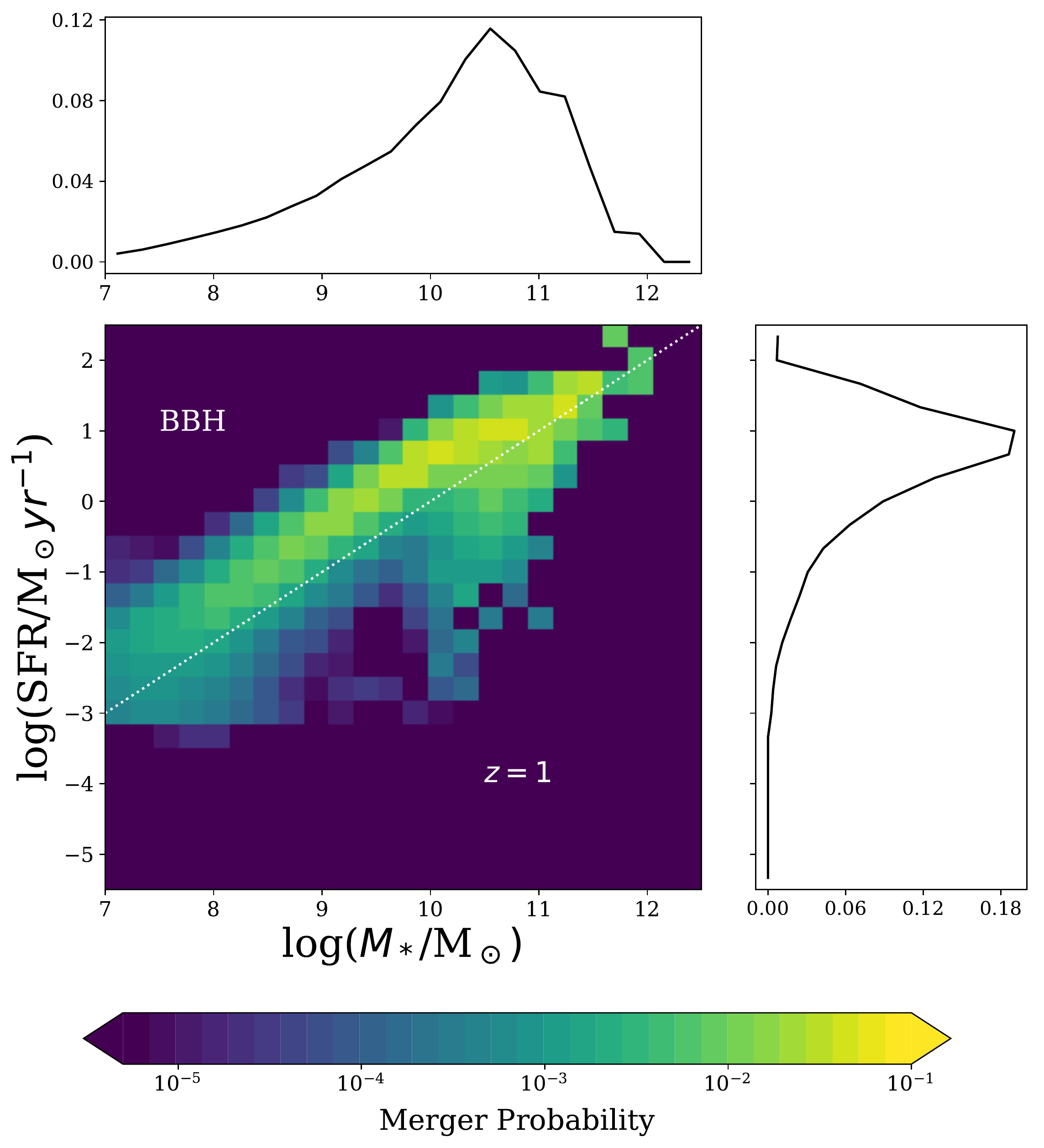}
\includegraphics[width=\columnwidth]{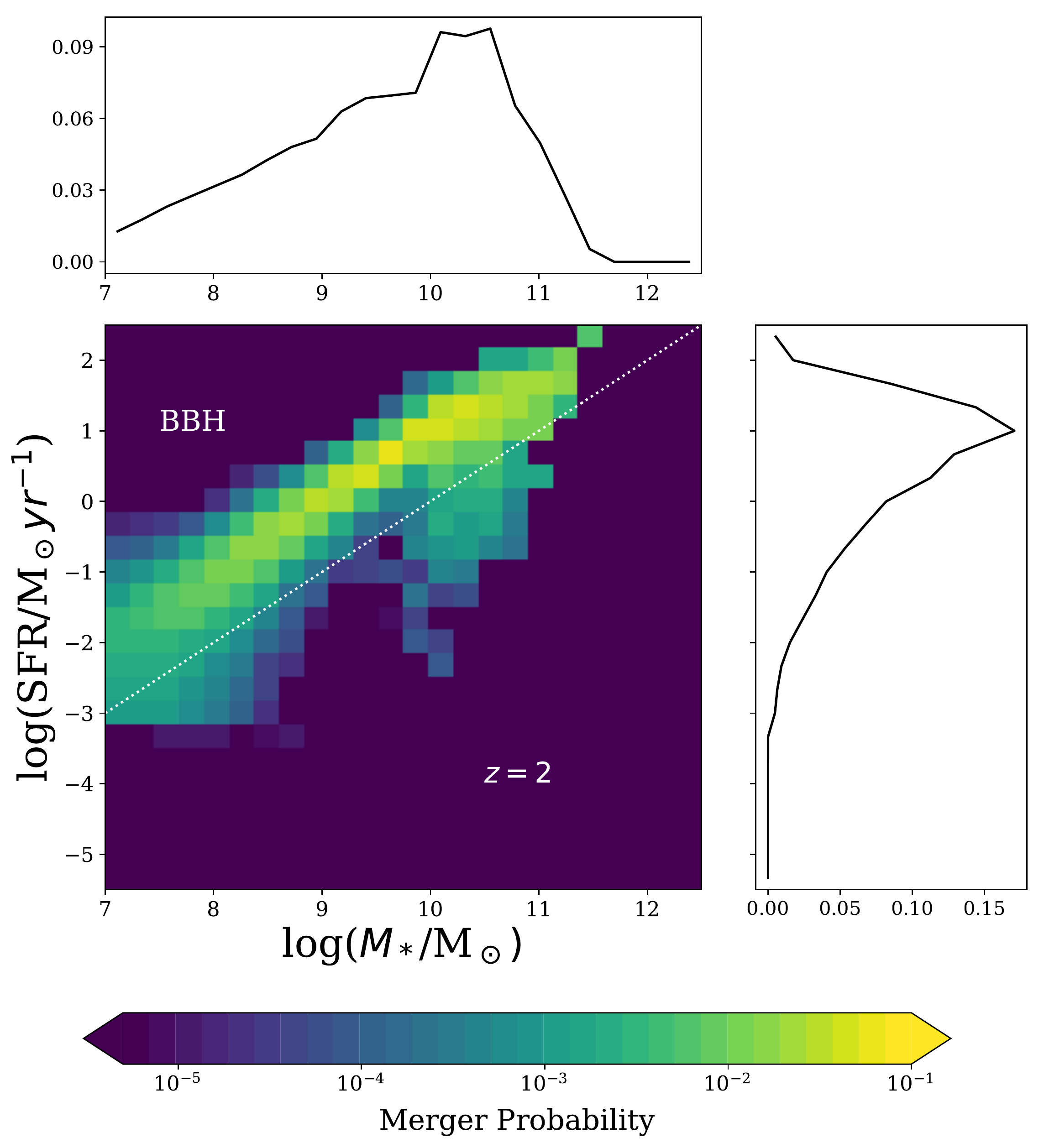}
\includegraphics[width=\columnwidth]{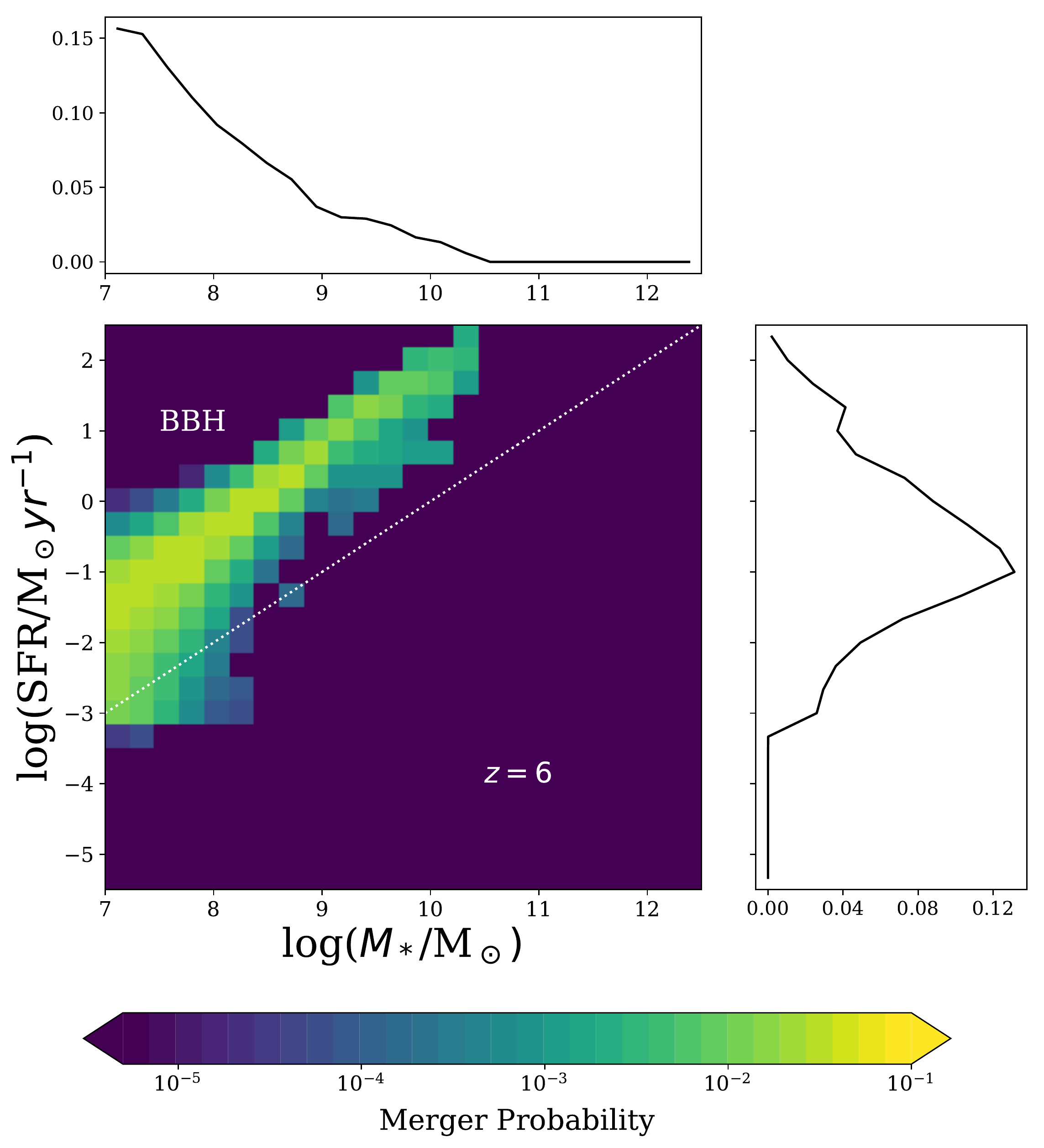}
\caption{Same as Figure~\ref{fig:DNS_prob_100Mpc}, but for BBHs.}
\label{fig:DBH_prob_100Mpc}
\end{figure*}

For each redshift interval, we compute the merger probability in a given bin of host's stellar mass and SFR $(M_{\ast{},i},\text{SFR}_{j})$, as
\begin{equation}\label{eq:prob}
    d^{X}_{i,j} = \frac{N_{{\rm GW}\,{}i,\,{}j}^X}{ N^X_{\rm tot,\,{}GW}},
\end{equation}
where $N_{{\rm GW}\,{}i,\,{}j}^X$ is the sum of all events of kind $X$ (where $X$ = BNSs, BHNSs or BBHs) occurring in galaxies with mass in the $i$th bin (between $M_{\ast{},i}-\Delta{}M_\ast/2$ and $M_{\ast{},i}+\Delta{}M_\ast/2$) and SFR in the $j$th bin (between ${\rm SFR}_i-\Delta{}{\rm SFR}/2$ and ${\rm SFR}_i+\Delta{}{\rm SFR}/2$) for a given time span.
The normalization $N^X_{\rm {tot,GW}}$ corresponds to the total number of mergers in all the galaxies of the simulated box, in a given time-span. As a consequence, summing over all the bins yields $\sum_{i,j} d^X_{i,j} = 1$.

The merger probability accounts for two different facts: i) the dependence of the number of compact-binary mergers on stellar mass and SFR of the host galaxy; ii) the fact that massive galaxies are less numerous than low-mass galaxies, manifested in the galaxy stellar mass function.

We note that the merger probability per a given host stellar mass and SFR can be interpreted as the probability of having a compact-binary merger for a given stellar mass and SFR. 
We stress that $d^{X}_{i,j}$ represents the merger probability in a given bin of SFRs and stellar masses: it must not be confused with the merger rate per galaxy (which is instead indicated as $n_{\rm GW}$ and is discussed in the previous Section~\ref{sec:results_fits}).

Figures~\ref{fig:DNS_prob_100Mpc},~\ref{fig:BHNS_prob_100Mpc} and ~\ref{fig:DBH_prob_100Mpc} show the merger probability of BNSs, BBHs and BHNSs, respectively, as a function of stellar mass and SFR at $z = 0.1, 1, 2$ and 6.
In each case we use a grid of $25\times25$ bins in the range of $\log(M_\ast{} / \Msun) \in{}[ 7.0,\,{}12.5]$ and $\log({\rm SFR} / \Msun \yr^{-1}) \in{}[ -5.5,\,{} 2.5]$. 
We find a strong correlation between merger probability, stellar mass and SFR. This correlation holds for both low and  high redshift.
In order to test the robustness of our results, we also computed the merger probability using the galaxy catalog from \eagle{}25 finding good agreement between the two simulated boxes. We present the comparison in Appendix~\ref{sec:appConv}.

Massive galaxies host progressively more compact-binary mergers as redshift decreases. This is apparent from the marginal histograms. To quantify the shift with redshift, we compute the median stellar mass from the marginal probability distributions. 
For merging BNSs, the median stellar masses of the hosts are $\log(M_\ast{}/M_{\odot}) = 10.4, 10.3, 10.2$, and 8.4 at $z = 0.1, 1, 2$, and 6, respectively. 
For merging BHNSs, the median values are $\log(M_\ast{}/M_{\odot}) = 10.7, 10.2, 9.7,$ and 8.0, at $z = 0.1, 1, 2$, and 6, respectively. 
Finally, for merging BBHs, the median stellar masses are $\log(M_\ast{}/M_{\odot}) = 10.9, 10.3, 9.8,$ and 7.8 at $z=0.1, 1, 2$, and 6, respectively. 
This result is expected since galaxies grow and become more massive with time.
The median of the stellar mass distribution for BBH hosts is shifted to a higher mass than for BNS, and BHNS hosts at $z=0.1$, in agreement with previous works \citep{Cao2018,Mapelli2018,Toffano2019}. This trend is reversed at high redshift, where the median is larger for BNS hosts.  

Finally, the host galaxy's SFR strongly correlates with $M_\ast{}$ at every considered redshift. This is a consequence of the mass -- SFR relation of galaxies \citep{Furlong2015}, as we already discussed in \cite{Artale2019}.
We also quantify the evolution of the SFR with redshift by computing median SFR from the marginal probability distributions.
For merging BNSs, the medians are $\log({\rm SFR}/M_\odot \yr^{-1}) = 0.07, 0.72, 1.0,$ and $0.02$ at $z=0.1, 1, 2,$ and 6, respectively.
For merging BHNSs, we get $\log({\rm SFR}/M_\odot \yr^{-1}) = 0.09, 0.65, 0.63,$ and $-0.46$ at $z=0.1, 1, 2,$ and 6, respectively.
For merging BBHs, we obtain $\log({\rm SFR}/M_\odot \yr^{-1}) = 0.12, 0.65, 0.64,$ and $-0.69$ at $z=0.1, 1, 2,$ and 6, respectively.
Our results show that the SFR medians reflect the evolution of the cosmic star formation rate in the Universe \citep{Madau2014}, regardless of the kind of merging compact object.

\subsection{Merger rate density evolution: early-type versus late-type host galaxies}\label{sec:ETLT}
 
Following \citet{Artale2019}, we estimate the contribution to the merger rate density from early-type (R$^{\rm ET}$) and late-type (R$^{\rm LT}$) galaxies at $z = 0.1, 1, 2$, and 6. We assume that galaxies with specific star formation rate (sSFR~=~SFR~/~$M_\ast{}$) lower than 10$^{-10}$ yr$^{-1}$ are early-type galaxies, while galaxies with sSFR~$\ge{}10^{-10}$ yr$^{-1}$ are late-type galaxies (see white dotted line in Fig.~\ref{fig:DNS_prob_100Mpc},~\ref{fig:BHNS_prob_100Mpc}, and \ref{fig:DBH_prob_100Mpc} as reference).

We note that there is no general consensus about the definition of late-type and early-type galaxies at different redshifts \citep[see e.g.,][]{Salim2007,Karim2011,Moustakas2013}. Our choice to distinguish between late-type and early-type galaxies based on a sSFR threshold (sSFR~ =~10$^{-10}$ yr$^{-1}$) must be regarded as one of the simplest possible assumptions. 
Different choices do not affect our main results significantly.

\begin{figure*}
\includegraphics[width=2.1\columnwidth]{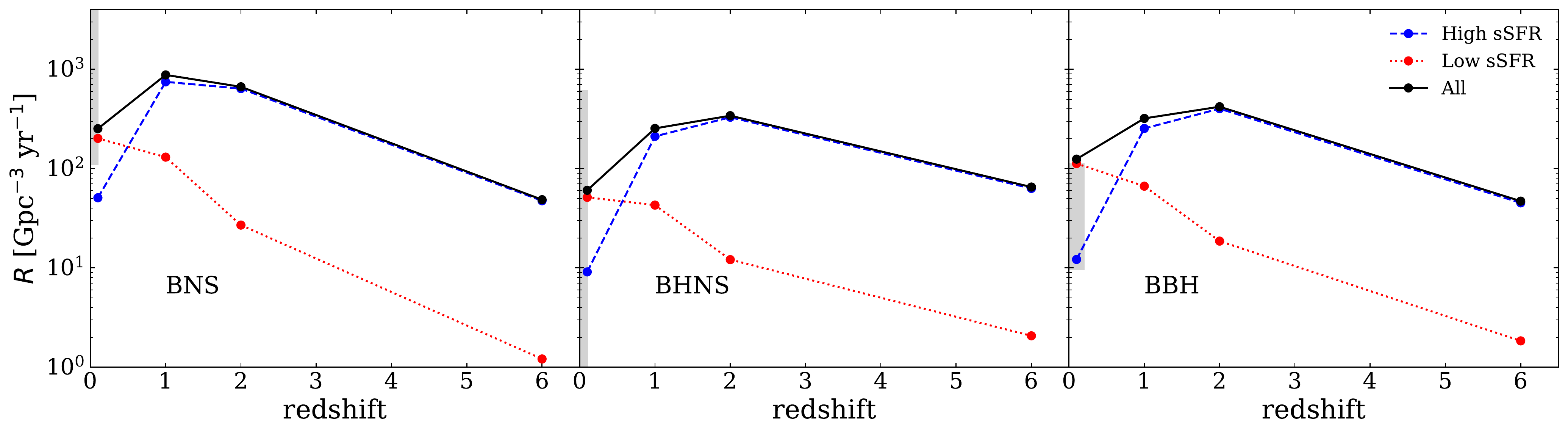}
\caption{Merger rate density of BNSs (left-hand panel), BHNSs (middle panel) and BBHs (right-hand panel) as a function of redshift. Black circles and solid lines: total merger rate density; red circles and dotted lines: contribution to the merger rate density of early-type galaxies; blue circles and dashed lines: contribution to the merger rate density of late-type galaxies.
 Gray regions: merger rate densities within 90~\% credible levers reported by the LIGO-Virgo collaboration \citep{AbbottO2}.}
\label{fig:MergerRateDensity_100Mpc}
\end{figure*}

Figure~\ref{fig:MergerRateDensity_100Mpc} shows the merger rate density of BNSs, BHNSs and BBHs as a function of redshift. We distinguish between the contribution to the merger rate density from early-type galaxies and that from late-type galaxies. 
Early-type galaxies give a larger contribution to the merger rate density at $z\leq{}0.1$ than late-type galaxies, in agreement with \citet{Artale2019}. The trend reverts at higher redshifts ($z\ge{}1$), where most mergers happen in late-type galaxies. This result springs from the combination of two effects. On the one hand, early-type galaxies become more and more common at low redshift, where they dominate the stellar mass budget (see e.g. \citealt{Moffett2016}). On the other hand, a large fraction of compact-binary mergers in the local Universe  are characterized by a long delay time (see e.g. \citealt{Mapelli2018,Mapelli2019}): these binary systems were born  several Gyr ago in galaxies with high sSFR, but merge at low redshift in galaxies with low sSFR and large stellar mass. 

\section{Conclusions}\label{sec:conclusions}
We have characterized the host galaxies of compact-binary mergers across cosmic time, from redshift $z=0$ to $z=6$, by means of population-synthesis simulations run with the code \mobse{} \citep{Mapelli2017,Giacobbo2018}, combined with galaxy catalogues from the \eagle{} cosmological simulations.

We find that there is a strong correlation between the merger rate per galaxy $n_{\rm GW}$ and the stellar mass $M_\ast{}$ of the host galaxy. This correlation holds at any redshifts considered here. We fit this correlation (Fit~1D) and we provide the best-fitting coefficients in Tables~\ref{table:Fits100MpcDNS}, \ref{table:Fits100MpcBHNS} and \ref{table:Fits100MpcDBH} for BNSs, BHNSs and BBHs, respectively. The slope of the $n_{\rm GW}-M_\ast$ correlation is steeper for BNSs than for both BHNSs and BBHs.

Moreover, we showed that including the SFR (Fit~2D) and the combination of SFR-metallicity (Fit~3D) improved the quality of our fit, especially in the latter case.
This demonstrates that the merger rate per galaxy depends not only on $M_\ast$ 
but also on SFR and $Z$. Hence, the merger rate per galaxy is maximum in galaxies with high mass, SFR and metallicity, both at low and high redshift (although galaxies are smaller at high $z$ and thus the typical mass of the host galaxies decreases with $z$).

We have shown that our results do not depend significantly on the numerical resolution and on the box size of the cosmological simulation.
Indeed, the correlations we find hold for both the \eagle{}100 (a 100 Mpc of side  comoving box simulation from the \eagle{} suite) and the \eagle{}25 (a 25 Mpc of side comoving box simulation from the \eagle{} suite).

By studying the merger probability as a function of $M_\ast$ and SFR (Figures~\ref{fig:DNS_prob_100Mpc}, \ref{fig:BHNS_prob_100Mpc} and \ref{fig:DBH_prob_100Mpc}), we find that this quantity evolves significantly with redshift for all considered compact binaries (BNSs, BHNSs and BBHs).
 The merger probability accounts for the fact that massive galaxies have a large merger rate per galaxy, together with the galaxy stellar mass function showing that massive galaxies are less numerous than low mass galaxies.
 
 At low redshift the merger probability shifts to higher stellar masses and to slightly lower SFR with respect to high redshift. This comes from the interplay between compact-binary delay time and cosmic evolution of the host galaxies.  The typical stellar masses of the host galaxies shift to higher values at low redshift, because galaxies grow with time due to cosmic assembly: a large portion of the stellar mass at low redshift is locked into relatively massive galaxies (see e.g. \citealt{Moffett2016}). Moreover, star formation decreases from $z\sim{}2$ to $z\sim{}0$. In addition, only binary compact objects with short delay time are able to merge  at high redshift, thus they tend to merge in the same galaxy where they formed \citep[see e.g.][]{Toffano2019}. In contrast, at low redshift we have both compact binaries that merge with short delay time and compact binaries that formed at high redshift and merge at low redshift with long delay time. Binaries with a long  delay time tend to merge in the most massive galaxies, where they ended up after cosmic galaxy assembly \citep{Mapelli2018}.

 Finally, we investigated the merger rate density of BNSs, BHNSs and BBHs (Figure~\ref{fig:MergerRateDensity_100Mpc}). We distinguish between the contribution to the merger rate density from late-type and early-type galaxies. At low redshift ($z\leq{}0.1$), the contribution to the merger rate from early-type galaxies is significantly larger than the one from late-type galaxies, while at higher redshift ($z\geq{}1$) most mergers occur in late-type galaxies.
   This trend is a consequence of the interplay between cosmic galaxy assembly and the delay time distribution of the compact-object mergers.
 
 Overall, our results predict a strong correlation between the merger rate and the properties (especially stellar mass and SFR) of host galaxies across cosmic time. Ongoing and future searches for electromagnetic counterparts of GW detections will probe this prediction, providing a fundamental test for compact-binary evolution models.

\section*{Acknowledgement}
We acknowledge the internal referee of LIGO-Virgo Giuseppe Greco for careful reading of the paper.
MCA and MM acknowledge financial support from the Austrian National Science Foundation through FWF stand-alone grant P31154-N27
``Unraveling merging neutron stars and black hole -- neutron star binaries with population synthesis simulations''.
MM and YB acknowledge financial support by the European Research Council for the ERC Consolidator grant DEMOBLACK, under contract no. 770017.
MS acknowledges funding from the European Union's Horizon 2020 research and innovation programme under the Marie-Sk\l{}odowska-Curie grant agreement No. 794393. MP acknowledges funding from the European Union's Horizon $2020$ research and innovation programme under the Marie Sk\l{}odowska-Curie grant agreement No. $664931$.
We acknowledge the Virgo Consortium for making the \eagle\ suite available. The \eagle\ simulations were performed using the DiRAC-2 facility
at Durham, managed by the ICC, and the PRACE facility Curie based in France at TGCC, CEA, Bruy\`{e}res-le-Ch\^{a}tel.

\bibliographystyle{mnras}

\bibliography{Artale_GW}

\IfFileExists{\jobname.bbl}{}
{\typeout{}
\typeout{****************************************************}
\typeout{****************************************************}
\typeout{** Please run "bibtex \jobname" to optain}
\typeout{** the bibliography and then re-run LaTeX}
\typeout{** twice to fix the references!}
\typeout{****************************************************}
\typeout{****************************************************}
\typeout{}
}

\appendix

\section{Comparison of \eagle{}25 and \eagle{}100}\label{sec:appConv}
In the main text, we have discussed the results we obtained from \eagle{}100, which represents the largest simulated box from the \eagle\ suite.
To test how robust our results are in terms of resolution and box size, we perform the same analysis using \eagle{}25. The \eagle{}25 simulation has a smaller box, but with a larger resolution than \eagle{}100 (gas particles are $\sim 8$ times smaller in \eagle{}25).  

\begin{figure*}
\includegraphics[width=\columnwidth]{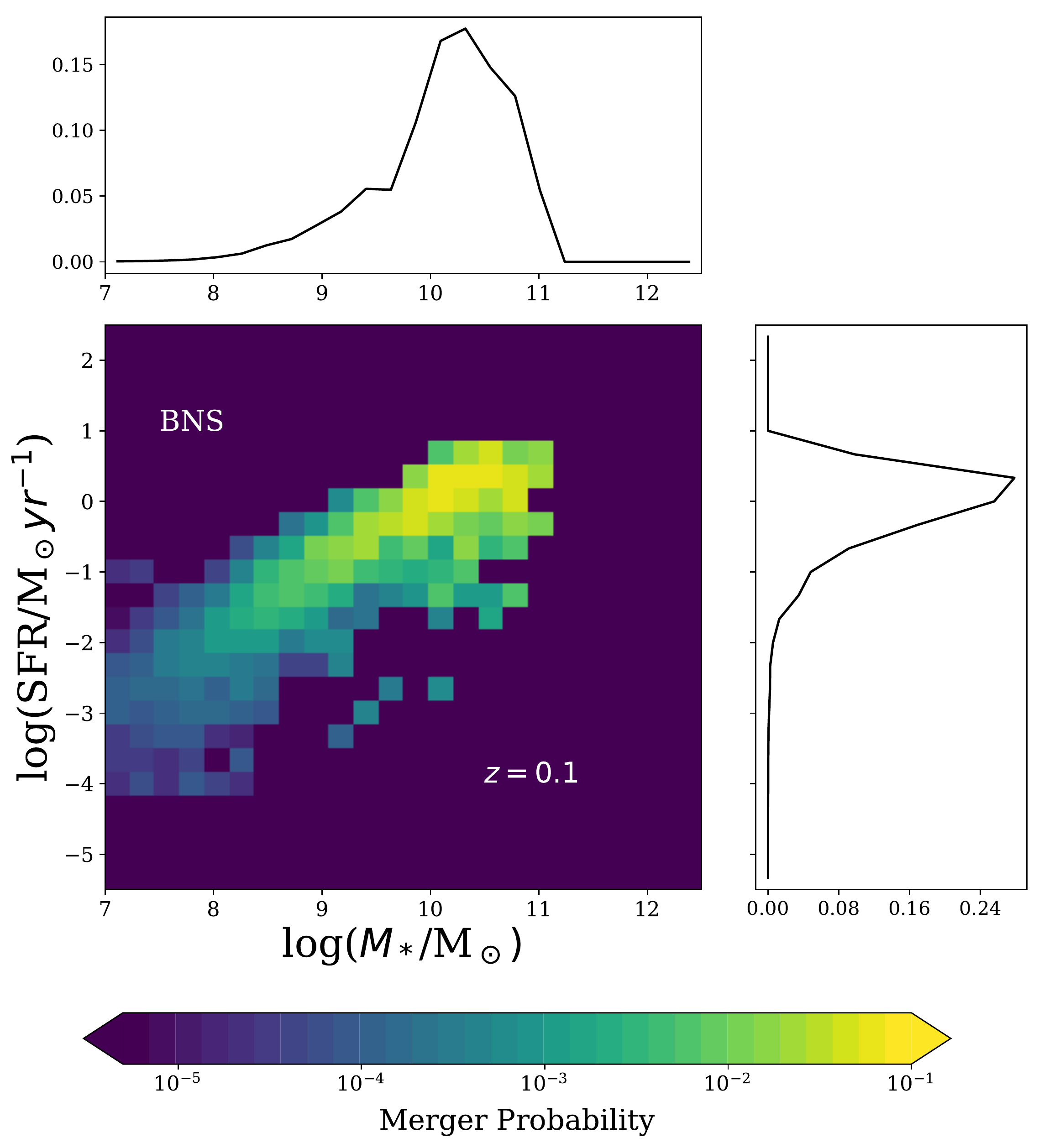}
\includegraphics[width=\columnwidth]{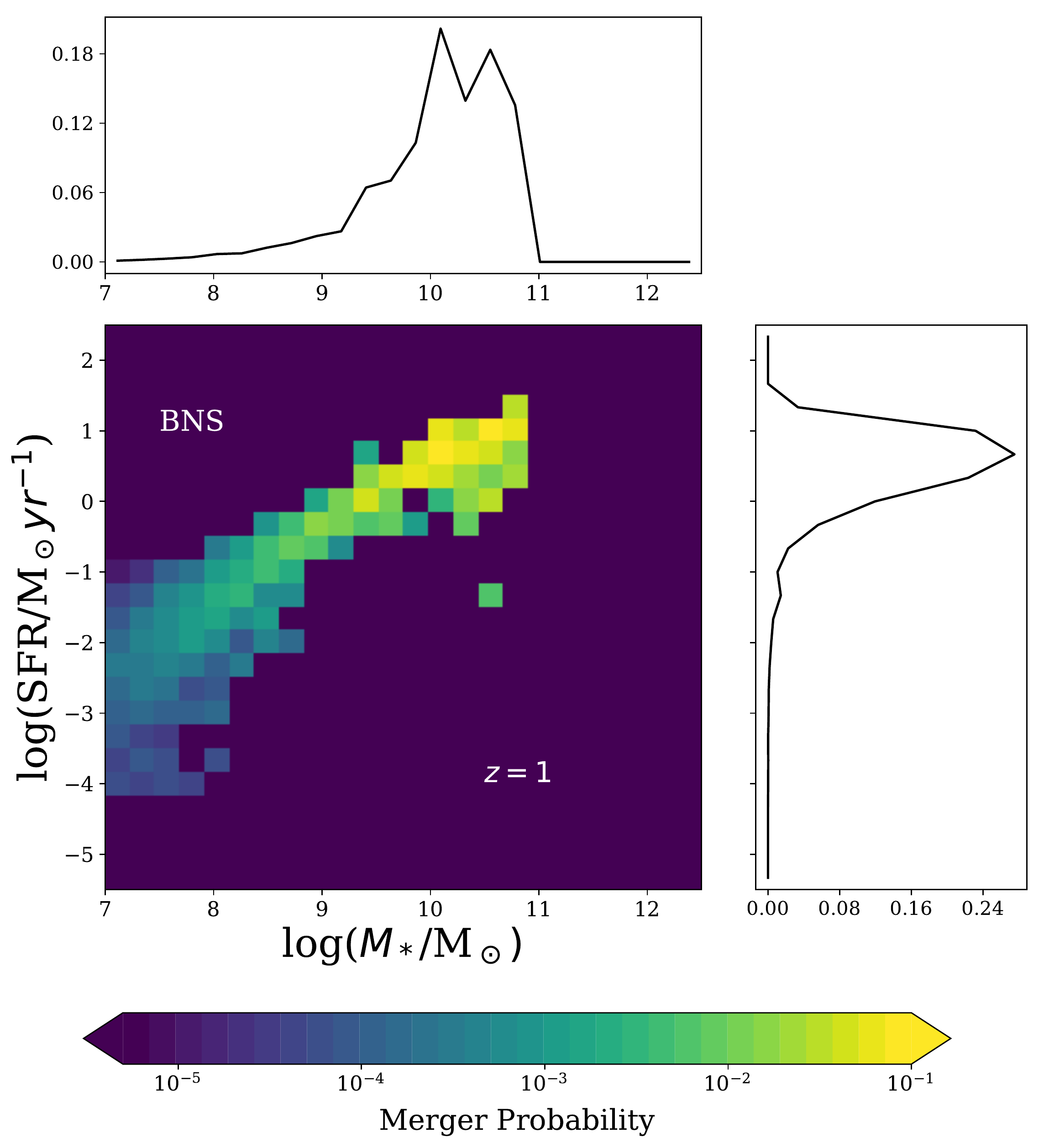}
\includegraphics[width=\columnwidth]{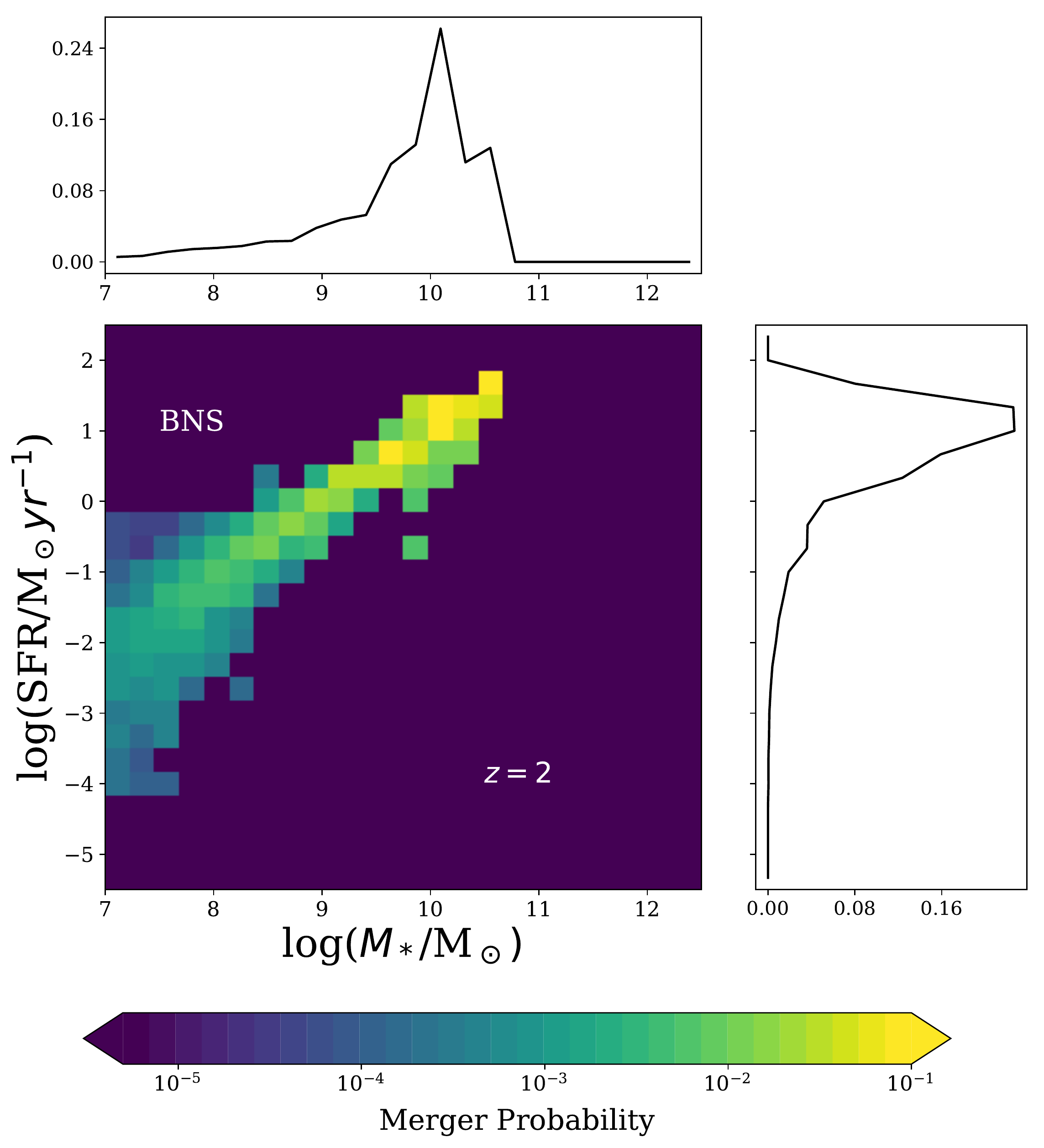}
\includegraphics[width=\columnwidth]{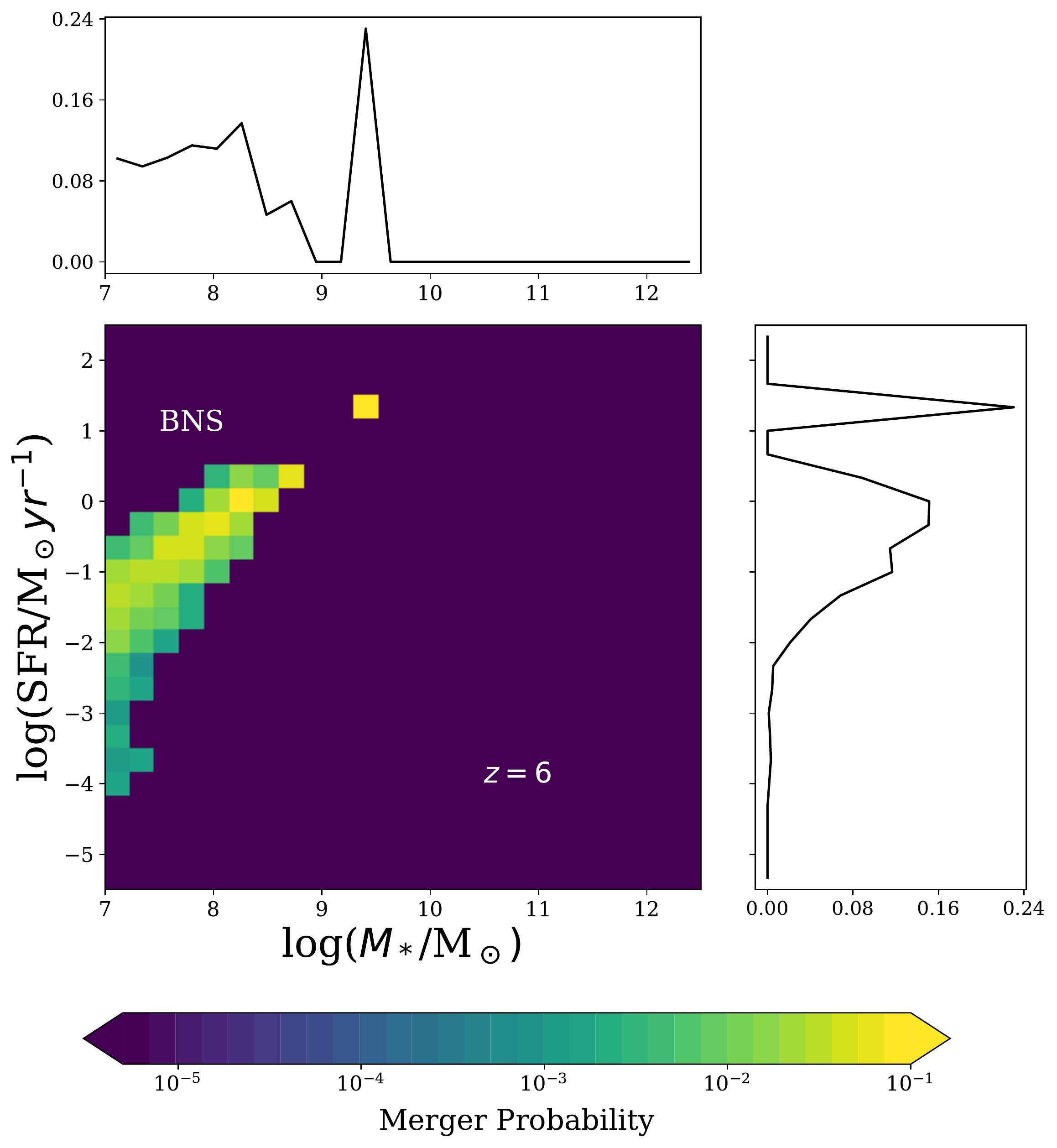}
\caption{Merger probability for BNSs at redshifts $z = 0.1, 1, 2$ and 6 as a function of the stellar mass and SFR for the \eagle{}25 galaxy catalogue.
Marginal histograms represent the distributions of SFR and $M_\ast{}$.}
\label{fig:DNS_prob_25Mpc}
\end{figure*}

\begin{figure*}
\includegraphics[width=\columnwidth]{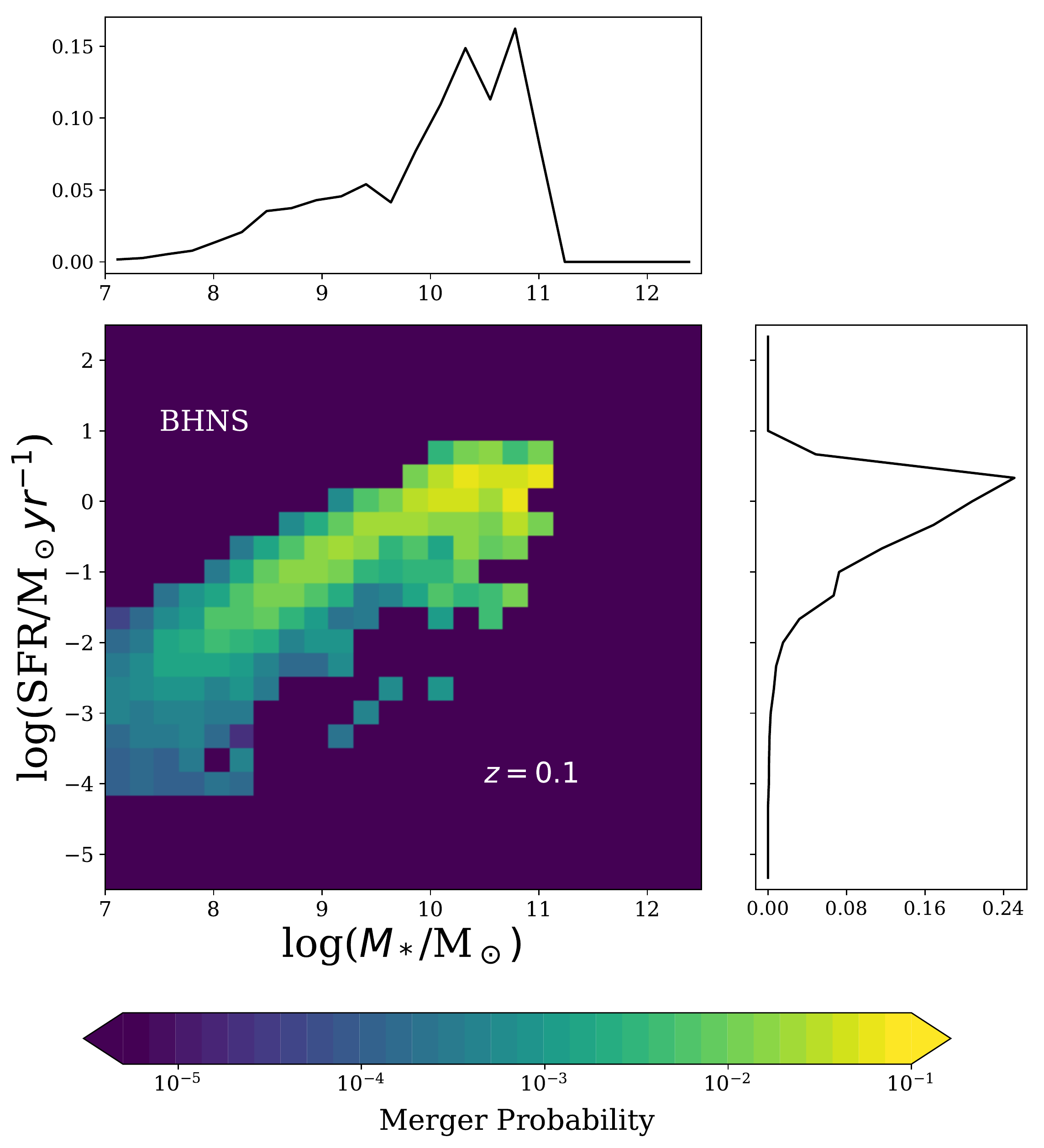}
\includegraphics[width=\columnwidth]{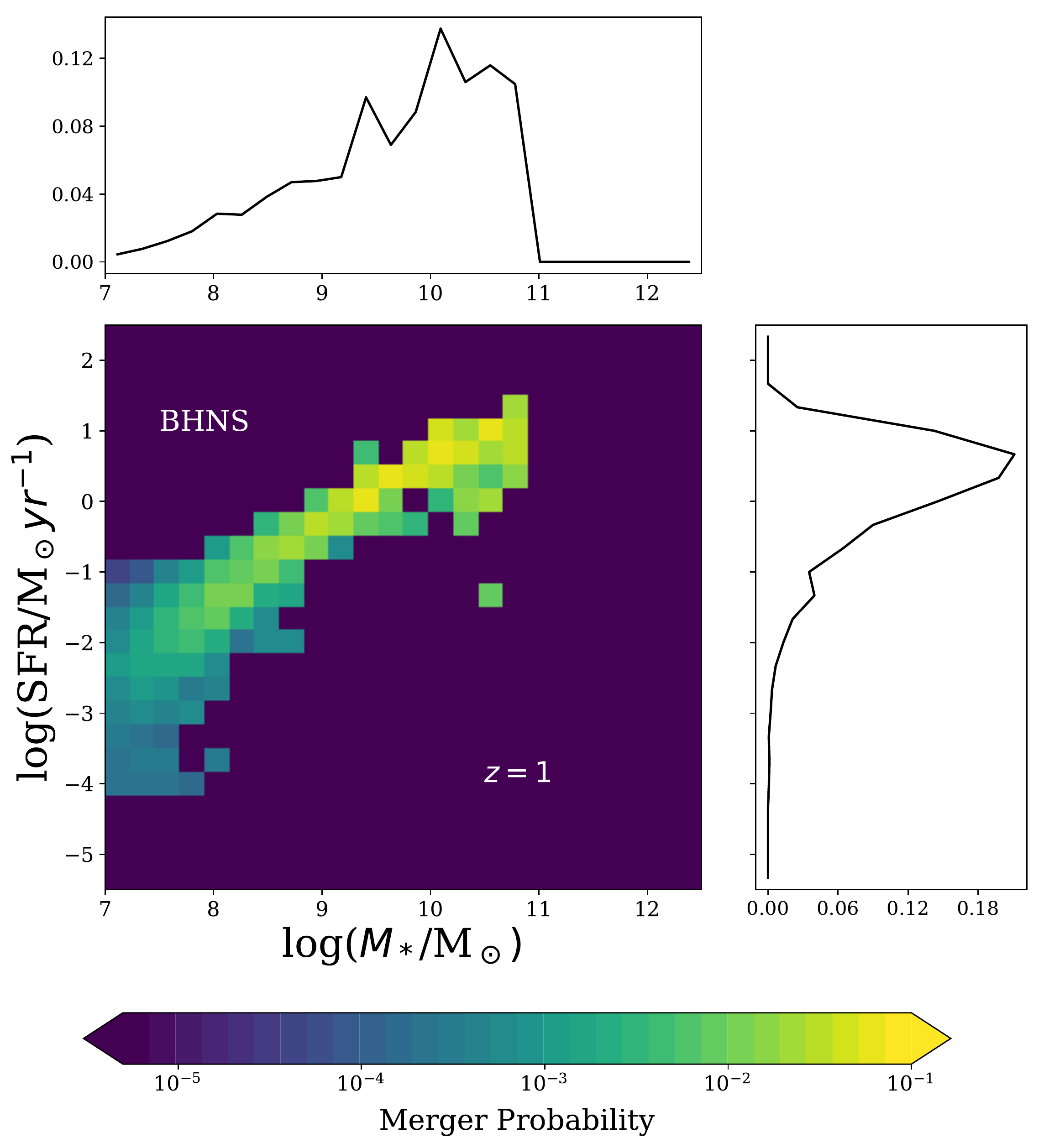}
\includegraphics[width=\columnwidth]{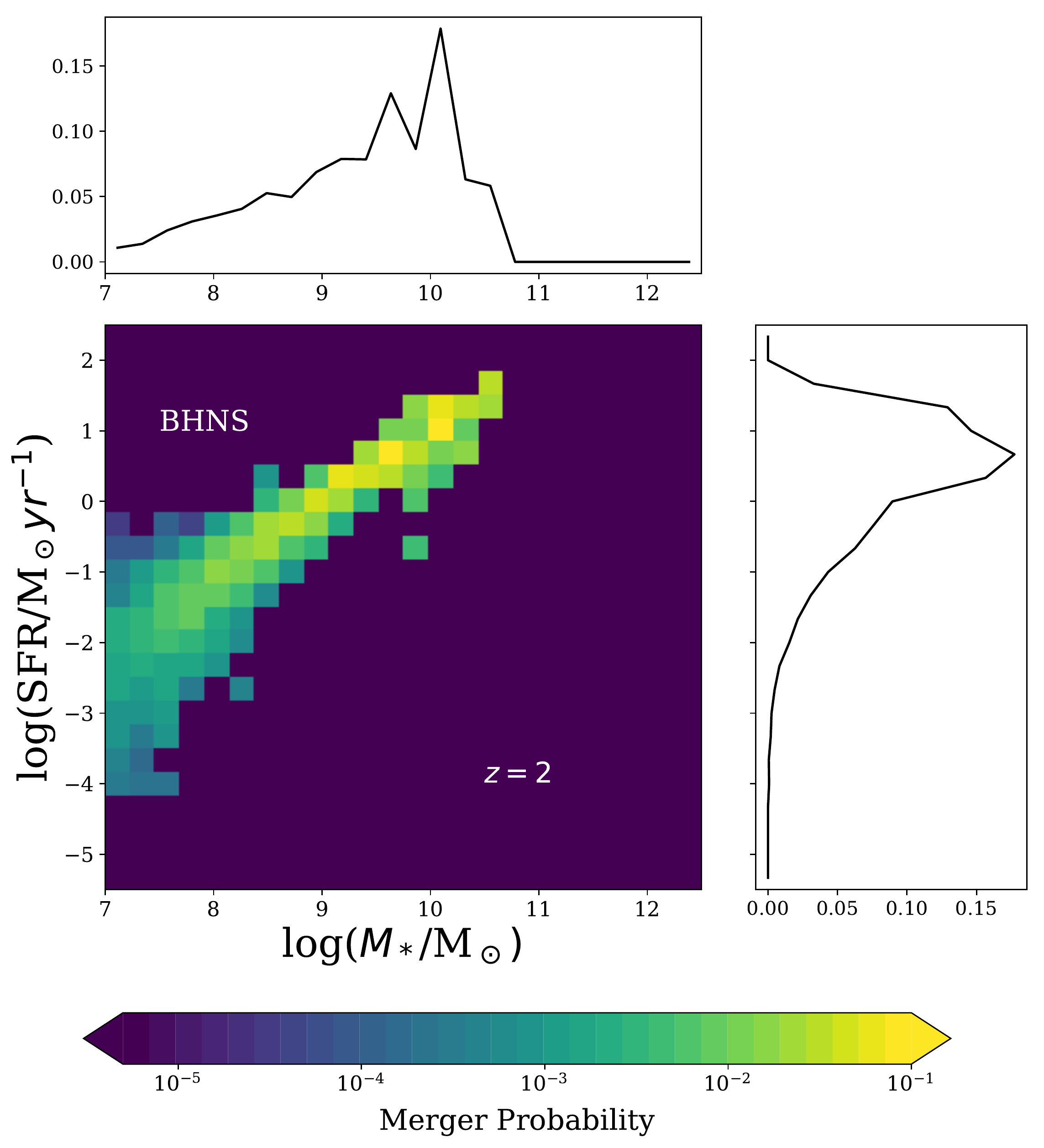}
\includegraphics[width=\columnwidth]{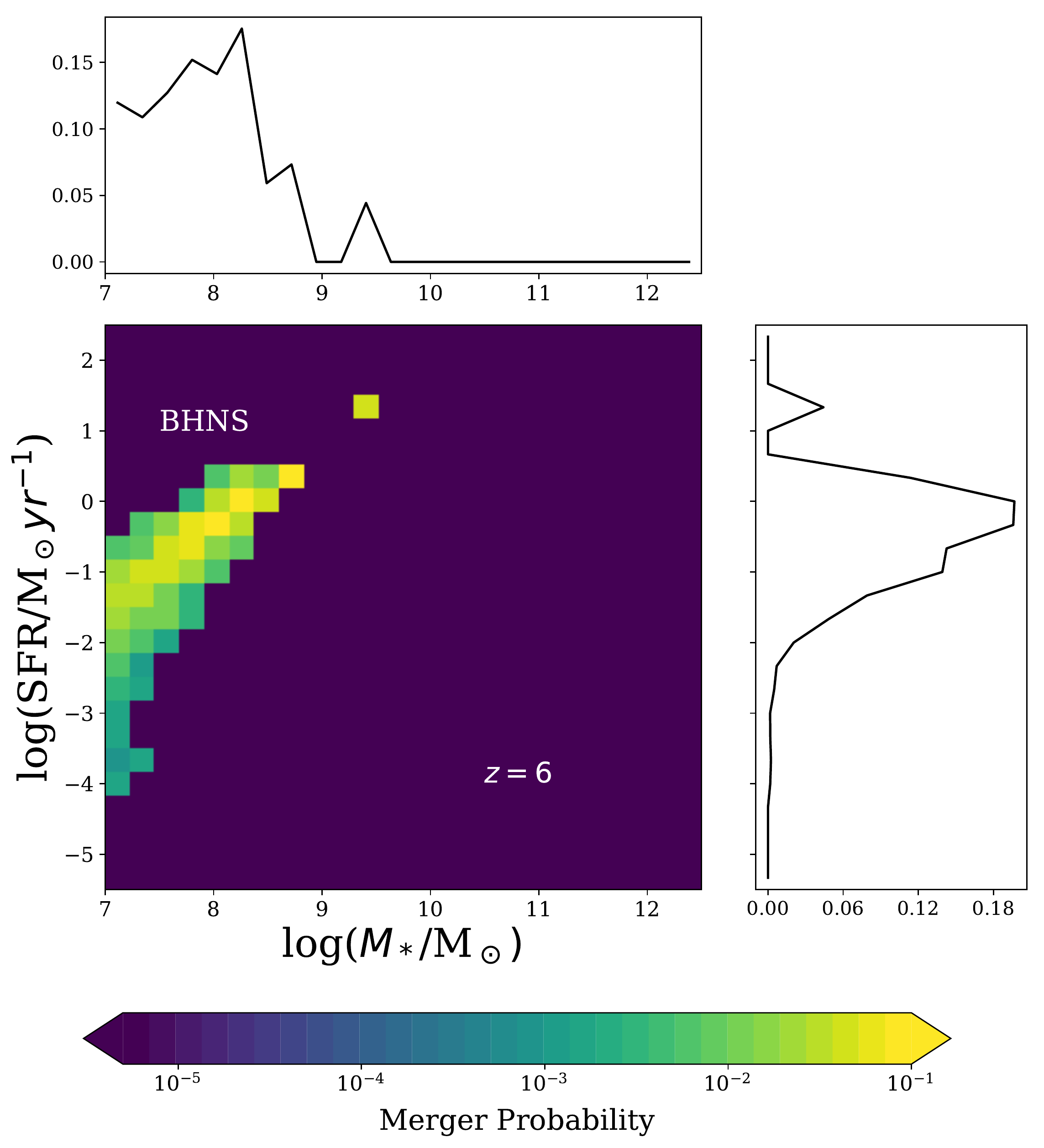}
\caption{Same as Figure~\ref{fig:DNS_prob_25Mpc} but for BHNSs.}
\label{fig:BHNS_prob_25Mpc}
\end{figure*}

\begin{figure*}
\includegraphics[width=\columnwidth]{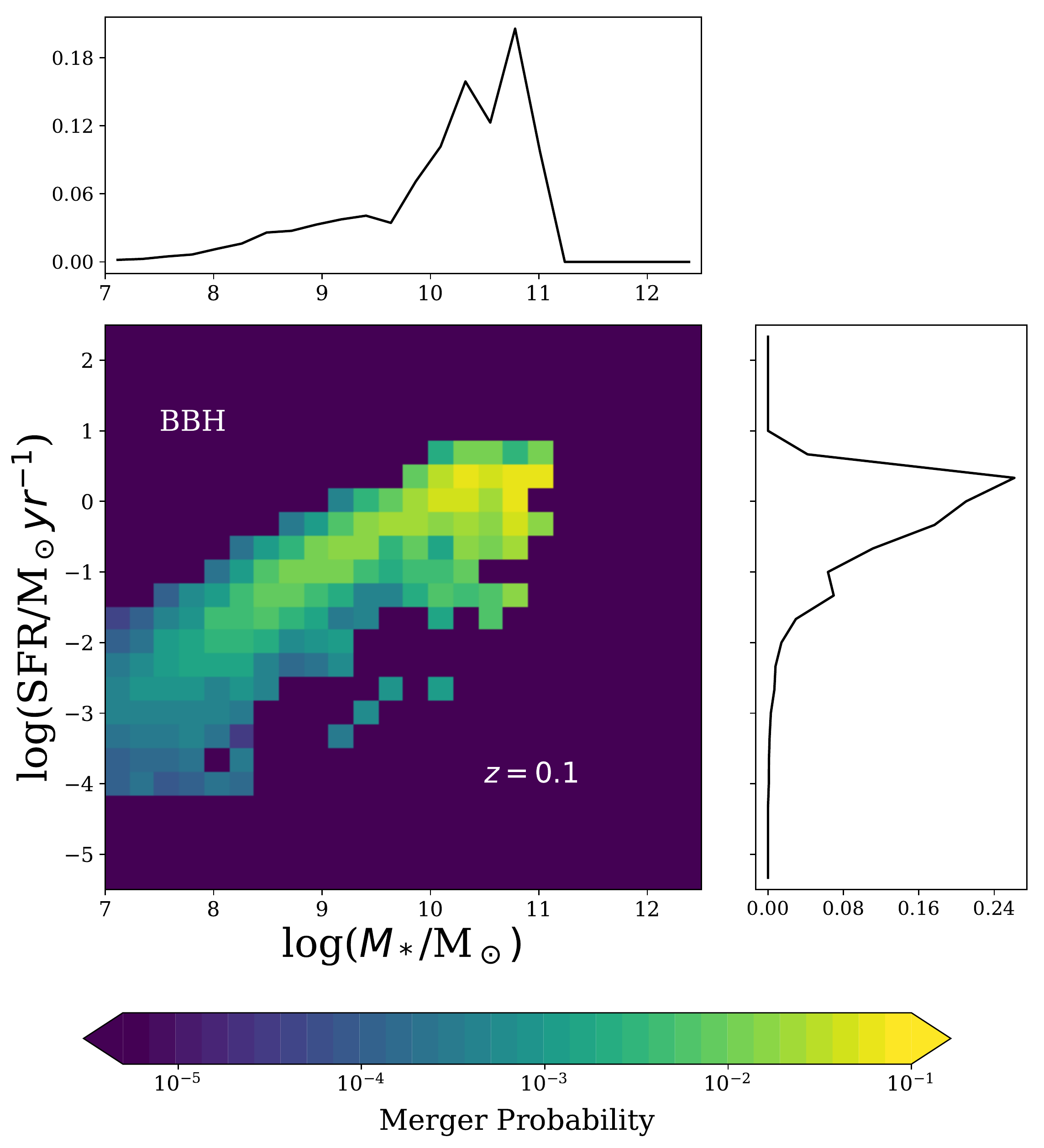}
\includegraphics[width=\columnwidth]{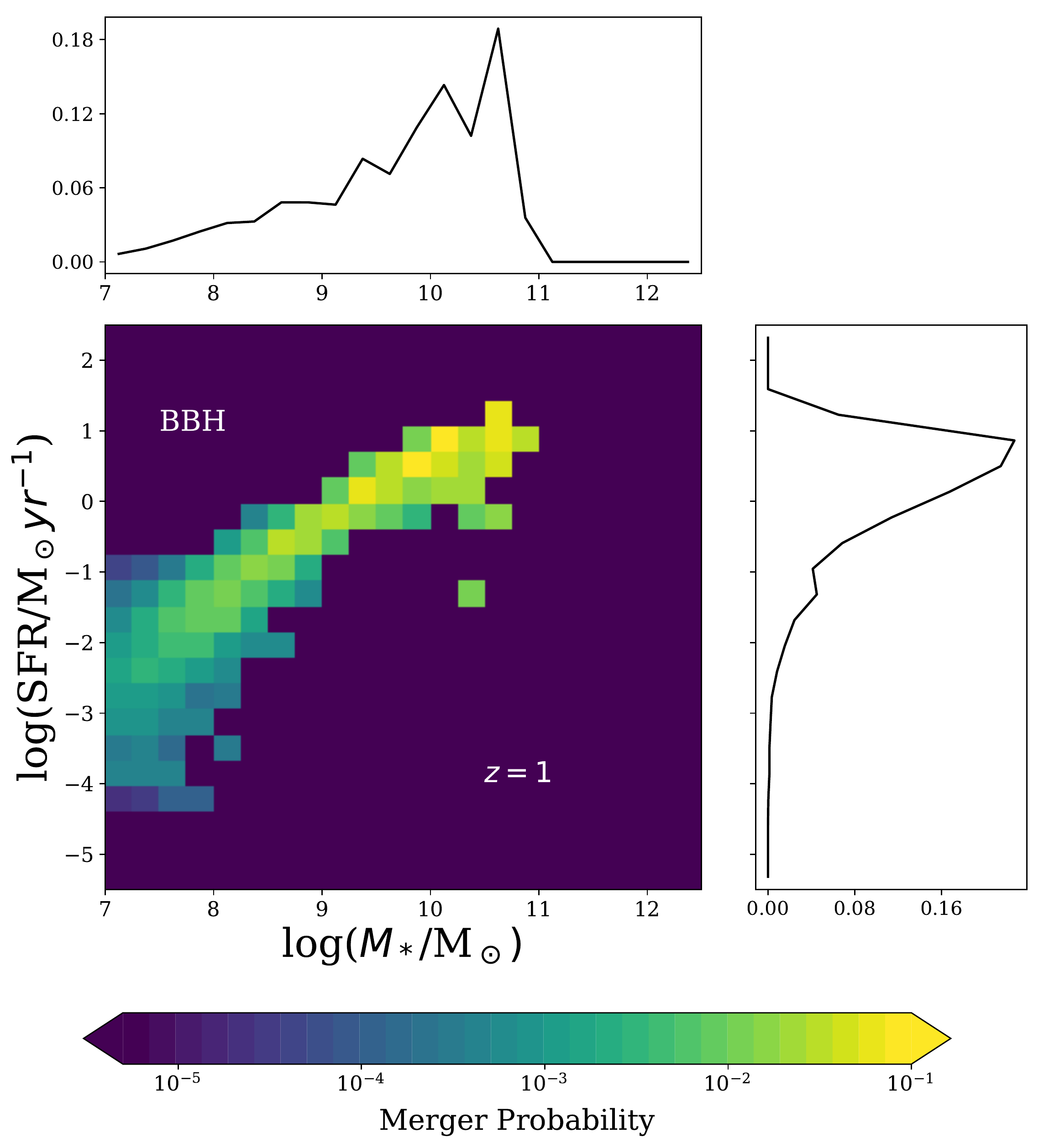}
\includegraphics[width=\columnwidth]{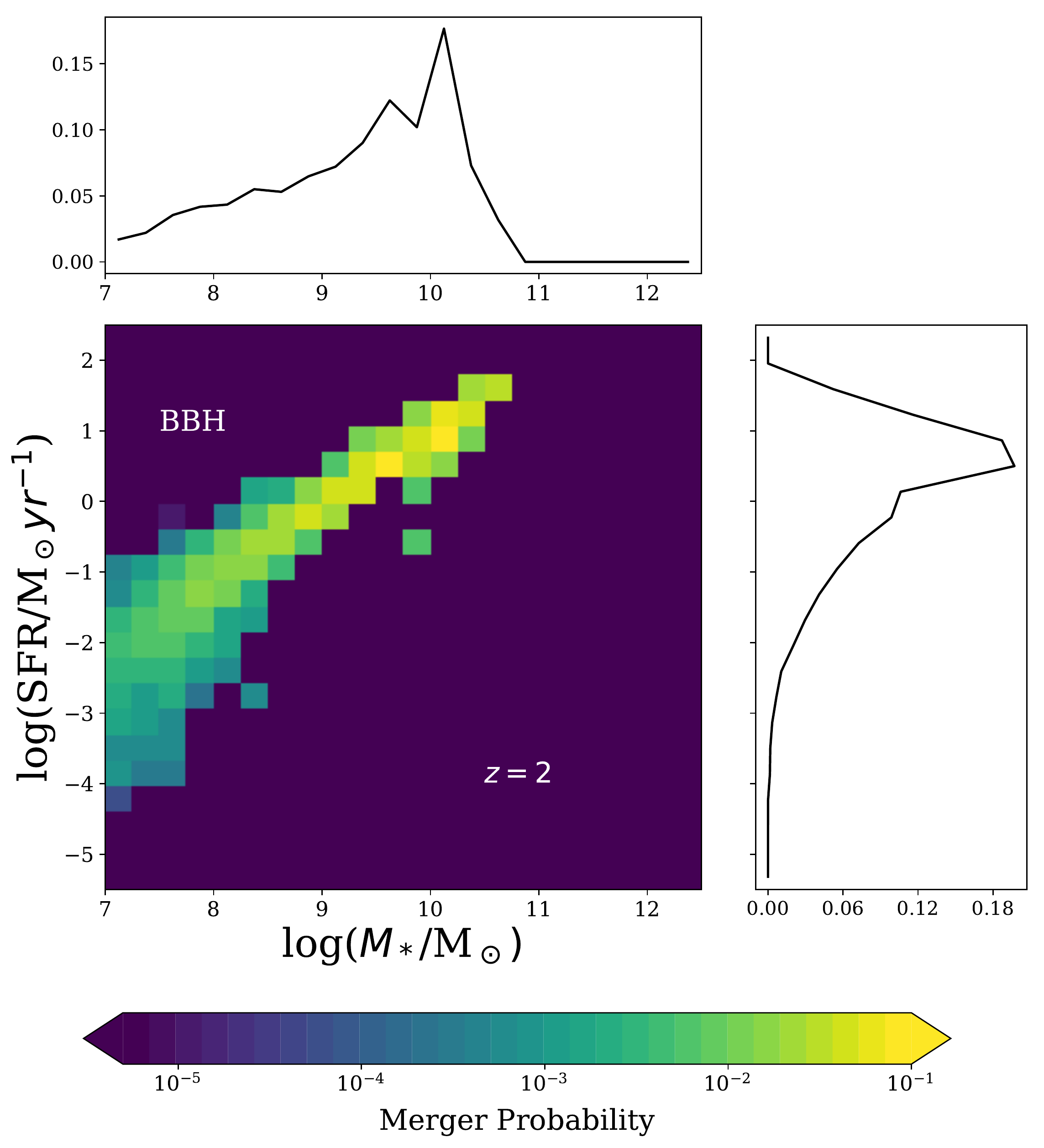}
\includegraphics[width=\columnwidth]{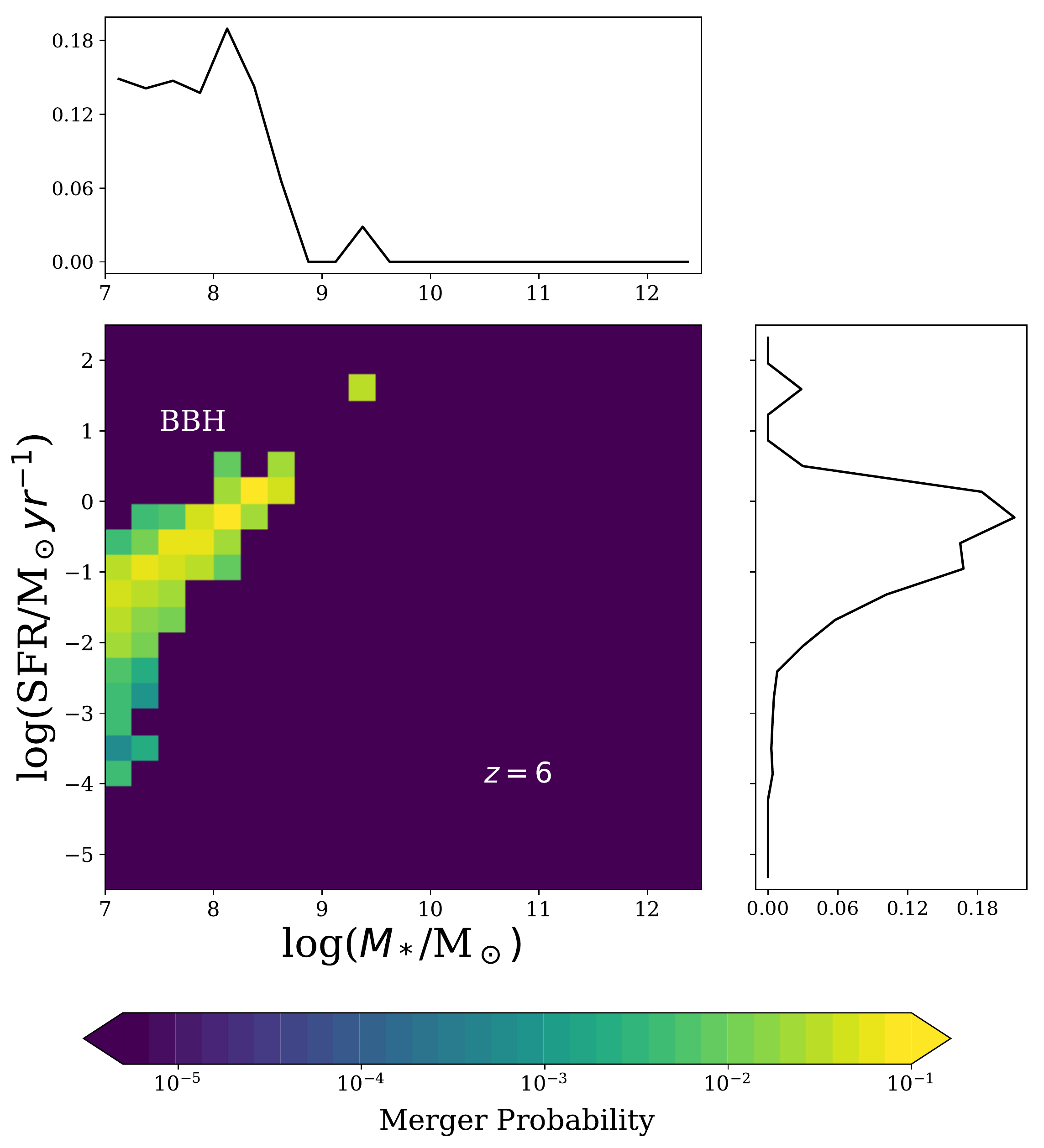}
\caption{Same as Figure~\ref{fig:DNS_prob_25Mpc} but for BBHs.}
\label{fig:DBH_prob_25Mpc}
\end{figure*}

Figures~\ref{fig:DNS_prob_25Mpc}, \ref{fig:BHNS_prob_25Mpc} and \ref{fig:DBH_prob_25Mpc} show the merger probability density for merging BNSs, BHNSs and BBHs, respectively, as a function of the stellar mass and SFR of the host galaxies. We bin the stellar mass and SFR using the same binning that we used for \eagle{}100.
 We find that the merger rates obtained with \eagle{}25 are similar to those derived from \eagle{}100. 
We note however that \eagle{}25 does not have massive clusters, due to the size of the simulated box. 
Tables~\ref{table:Fits25MpcDNS},  \ref{table:Fits25MpcBHNS} and \ref{table:Fits25MpcDBH} show the results of the fits proposed in  Sec.~\ref{sec:results_fits}. 
Our results show that the coefficients of Fit 1D for \eagle{}100 and \eagle{}25 are in good agreement, with percentage errors smaller than $\sim 8\%$.
This again indicates that the stellar mass of host galaxies of merging compact objects is a crucial property to trace their merger rates.
For Fits 2D and 3D, we also find agreement between the two galaxy catalogues, with only few cases showing large discrepancies with percentage errors of the coefficients standing above $\sim 20\%$. It is worth mentioning that only the coefficients attached to SFR and metallicity are concerned.
This is expected, because SFR and metallicity are more affected by simulation resolution and box size, as they strongly depend on sub-grid prescriptions
\citep[see e.g.,][]{Furlong2015,DeRossi2017}.


\begin{table*}
\caption{Same as Table~\ref{table:Fits100MpcDNS} for BNSs, but using \eagle{}25.}\label{table:Fits25MpcDNS}
\begin{tabular}{cc|c|c|c|c|l}
\cline{3-6}
& & \multicolumn{4}{ c| }{BNSs} \\ \cline{3-6}
& & $z = 0.1$ & $z=1$ & $z=2$ & $z=6$ \\ \cline{1-6}
\multicolumn{1}{ |c  }{\multirow{2}{*}{Fit~1D} } &
\multicolumn{1}{ |c| }{$\alpha_{1}$} & $1.073   \pm  0.005 $  & $1.083   \pm  0.005 $  & $0.992   \pm  0.006 $  & $0.999   \pm  0.017 $  &     \\ \cline{2-6}
\multicolumn{1}{ |c  }{}                        &
\multicolumn{1}{ |c| }{ $\alpha_{2}$} & $-6.512    \pm  0.046 $  & $-6.165   \pm  0.039 $   & $-5.099    \pm  0.044 $   & $-4.886     \pm  0.130$   &     \\ \cline{1-6}
\multicolumn{1}{ |c  }{\multirow{2}{*}{Fit~2D} } &
\multicolumn{1}{ |c| }{$\beta_{1}$} & $0.865    \pm  0.010  $  & $1.025    \pm   0.010  $  & $0.960     \pm  0.010  $  & $1.021     \pm  0.025$   &  \\ \cline{2-6}
\multicolumn{1}{ |c  }{}                        &
\multicolumn{1}{ |c| }{$\beta_{2}$} & $0.217    \pm  0.009  $  & $0.053     \pm  0.008  $   & $0.026    \pm  0.007  $   & $-0.016    \pm  0.013 $   &  \\ \cline{2-6}
\multicolumn{1}{ |c  }{}                        &
\multicolumn{1}{ |c| }{$\beta_{3}$} & $-4.392    \pm  0.096 $   & $-5.618      \pm   0.090  $   & $-4.816    \pm  0.089  $   & $-5.073    \pm   0.199$   &  \\ \cline{1-6}
\multicolumn{1}{ |c  }{\multirow{2}{*}{Fit~3D} } &
\multicolumn{1}{ |c| }{$\gamma_{1}$} & $0.852    \pm  0.015  $   & $1.096   \pm  0.015   $   & $1.076    \pm 0.016   $   & $1.054    \pm    0.037$   &  \\ \cline{2-6}
\multicolumn{1}{ |c  }{}                        &
\multicolumn{1}{ |c| }{$\gamma_{2}$} &  $0.225    \pm  0.011  $  & $0.014    \pm  0.010  $  & $-0.029   \pm   0.009  $   & $-0.033    \pm    0.020$   & \\ \cline{2-6}
\multicolumn{1}{ |c  }{}                        &
\multicolumn{1}{ |c| }{$\gamma_{3}$} & $0.031   \pm  0.027  $   & $-0.136   \pm  0.022   $   & $-0.235   \pm  0.025 $   & $-0.067   \pm    0.057 $   & \\ \cline{2-6}
\multicolumn{1}{ |c  }{}                        &
\multicolumn{1}{ |c| }{$\gamma_{4}$} & $-4.214    \pm   0.182 $   & $-6.553    \pm  0.179   $  & $-6.386    \pm   0.189 $   & $-5.524    \pm   0.434 $  & \\ \cline{1-6}
\end{tabular}
\end{table*}
\begin{table*}
\caption{Same as Table~\ref{table:Fits100MpcBHNS} for BHNSs but using  \eagle{}25.}\label{table:Fits25MpcBHNS}
\begin{tabular}{cc|c|c|c|c|l}
\cline{3-6}
& & \multicolumn{4}{ c| }{BHNSs} \\ \cline{3-6}
& & $z = 0.1$ & $z=1$ & $z=2$ & $z=6$ \\ \cline{1-6}
\multicolumn{1}{ |c  }{\multirow{2}{*}{Fit~1D} } &
\multicolumn{1}{ |c| }{$\alpha_{1}$} & $0.820   \pm  0.007 $   & $0.850   \pm  0.008 $   & $0.915   \pm   0.006$    & $1.012   \pm   0.015$   &     \\ \cline{2-6}
\multicolumn{1}{ |c  }{}                        &
\multicolumn{1}{ |c| }{ $\alpha_{2}$} & $-4.510   \pm  0.058 $   & $-4.149   \pm  0.062 $  & $-4.357   \pm   0.047$   & $-4.726   \pm   0.114$    &     \\ \cline{1-6}
\multicolumn{1}{ |c  }{\multirow{2}{*}{Fit~2D} } &
\multicolumn{1}{ |c| }{$\beta_{1}$} & $0.604   \pm  0.013 $   & $0.719   \pm  0.016 $  & $0.860   \pm   0.011$   &  $0.976   \pm   0.022$   &  \\ \cline{2-6}
\multicolumn{1}{ |c  }{}                        &
\multicolumn{1}{ |c| }{$\beta_{2}$} & $0.224   \pm  0.012 $   & $0.120   \pm  0.012 $   & $0.046   \pm  0.008 $    &  $0.026   \pm  0.011 $  &  \\ \cline{2-6}
\multicolumn{1}{ |c  }{}                        &
\multicolumn{1}{ |c| }{$\beta_{3}$} & $-2.318   \pm   0.127$   & $-2.907   \pm   0.142$   & $-3.863   \pm   0.094$   & $-4.423   \pm   0.174$   &  \\ \cline{1-6}
\multicolumn{1}{ |c  }{\multirow{2}{*}{Fit~3D} } &
\multicolumn{1}{ |c| }{$\gamma_{1}$} & $1.011   \pm  0.016 $   & $1.171   \pm  0.020 $  & $1.154   \pm  0.0143 $  & $0.990   \pm   0.033$  &  \\ \cline{2-6}
\multicolumn{1}{ |c  }{}                        &
\multicolumn{1}{ |c| }{$\gamma_{2}$} &  $-0.031   \pm 0.012  $   & $-0.123   \pm   0.013$  & $-0.096   \pm  0.008 $    & $0.019   \pm   0.017$  & \\ \cline{2-6}
\multicolumn{1}{ |c  }{}                        &
\multicolumn{1}{ |c| }{$\gamma_{3}$} & $-0.994   \pm 0.028  $   & $-0.874   \pm  0.029 $   & $-0.598   \pm  0.023 $   & $-0.029   \pm   0.050$   & \\ \cline{2-6}
\multicolumn{1}{ |c  }{}                        &
\multicolumn{1}{ |c| }{$\gamma_{4}$} & $-8.036   \pm  0.193 $  & $-8.863   \pm   0.230$  &  $-7.862   \pm   0.171$  & $-4.619   \pm  0.379 $   & \\ \cline{1-6}
\end{tabular}
\end{table*}
\begin{table*}
\caption{Same as Table~\ref{table:Fits100MpcDBH} for BBHs but using  \eagle{}25.}\label{table:Fits25MpcDBH}
\begin{tabular}{cc|c|c|c|c|l}
\cline{3-6}
& & \multicolumn{4}{ c| }{BBHs} \\ \cline{3-6}
& & $z = 0.1$ & $z=1$ & $z=2$ & $z=6$ \\ \cline{1-6}
\multicolumn{1}{ |c  }{\multirow{2}{*}{Fit~1D} } &
\multicolumn{1}{ |c| }{$\alpha_{1}$} & $0.792 \pm  0.006 $  & $0.792 \pm  0.008  $ & $0.838 \pm  0.007  $ & $0.918 \pm  0.019  $ &     \\ \cline{2-6}
\multicolumn{1}{ |c  }{}                        &
\multicolumn{1}{ |c| }{ $\alpha_{2}$} & $-4.059 \pm  0.046  $ & $-3.504 \pm   0.061  $ & $-3.537 \pm  0.053  $ & $-4.086   \pm  0.139 $ &     \\ \cline{1-6}
\multicolumn{1}{ |c  }{\multirow{2}{*}{Fit~2D} } &
\multicolumn{1}{ |c| }{$\beta_{1}$} & $0.700 \pm   0.011  $ & $0.728 \pm   0.016  $ & $0.850 \pm  0.012  $ & $0.966  \pm  0.027  $ &  \\ \cline{2-6}
\multicolumn{1}{ |c  }{}                        &
\multicolumn{1}{ |c| }{$\beta_{2}$} & $0.095  \pm  0.010  $ & $0.058 \pm   0.013  $ & $-0.010 \pm   0.009  $ & $-0.034 \pm  0.014  $ &  \\ \cline{2-6}
\multicolumn{1}{ |c  }{}                        &
\multicolumn{1}{ |c| }{$\beta_{3}$} & $-3.124 \pm   0.108  $ & $-2.905 \pm  0.145   $ & $-3.639  \pm   0.108 $ & $-4.482  \pm  0.212 $ &  \\ \cline{1-6}
\multicolumn{1}{ |c  }{\multirow{2}{*}{Fit~3D} } &
\multicolumn{1}{ |c| }{$\gamma_{1}$} & $0.993 \pm  0.015   $ & $1.167 \pm  0.020  $ & $1.234 \pm  0.015  $ & $1.096   \pm 0.039 $ &  \\ \cline{2-6}
\multicolumn{1}{ |c  }{}                        &
\multicolumn{1}{ |c| }{$\gamma_{2}$} & $-0.090 \pm  0.011  $ & $-0.176 \pm  0.013  $ & $-0.196 \pm  0.009  $ & $-0.104  \pm  0.020 $ & \\ \cline{2-6}
\multicolumn{1}{ |c  }{}                        &
\multicolumn{1}{ |c| }{$\gamma_{3}$} & $-0.699 \pm  0.027   $ & $-0.851  \pm  0.031 $ & $-0.782 \pm  0.024  $ & $-0.269  \pm 0.059 $ & \\ \cline{2-6}
\multicolumn{1}{ |c  }{}                        &
\multicolumn{1}{ |c| }{$\gamma_{4}$} & $-7.211  \pm  0.182   $ & $-8.684 \pm  0.238  $ & $-8.867 \pm  0.180  $ & $-6.305  \pm 0.451  $ & \\ \cline{1-6}
\end{tabular}
\end{table*}

\section{Change of slope in the $n_{\rm GW}$ -- $M_\ast{}$ correlation}\label{sec:galprop_mergRates} 
From a visual inspection of Figure~\ref{fig:nGW_Ms_100Mpc}, it appears that there is a subtle change of the slope of BBH hosts galaxies at $\log(M_\ast{}/M_\odot)\sim 10.5$ for redshift $z=0.1,$ 1 and 2. This trend does not appear in BNS hosts, and is less clear for BHNS hosts.

We measure how significant the slope change is, by using two different (but related) methods.
First, based on Fit~1D from Sec.~\ref{sec:results_fits}, we perform an ordinary least squares regression for the two following models (where $M_\ast{}$ is taken in units of $M_\odot$, $\theta(x)$ is the Heaviside step function, and $\epsilon$ represents an independent, identically distributed noise term):
\begin{equation}
\label{constantslope}
\mathtt{BBH} = a \log(M_\ast{}) + b + \epsilon
\end{equation}
and
\begin{equation}
\label{slopechange}
\mathtt{BBH} = a \log(M_\ast{}) + b + c\theta(\log(M_\ast{}) - 10.5) \log(M_\ast{}) + \epsilon.
\end{equation}
For the latter fit we obtain that the $c$ coefficient is significant to $p < 2 \cdot 10^{-16}$, corresponding to a value of the $t$ statistic of $~59$.
We also use the Bayesian Information Criterion (BIC) to compare the goodness of the fit for the two models. The BIC can be interpreted as a penalized log-likelihood of the model, where the penalty term takes into account the number of free parameters \citep[][]{ando2010bayesian}.
The more negative the BIC is, the better is the model. We find BIC $=  -46089$ for the first model (without slope change), while BIC $= -49479$ for the second model with slope change, which is unsurprisingly aligned with our results based on the adjusted R-squared.

The slope change of the host galaxies of BBH mergers arises from the interplay between the MZR and the strong dependence on metallicity of BBH progenitors. In fact, the MZR is steep for galaxies with $M_\ast{} \sim 10^{8.5} - 10^{10.5} \Msun$, but has a turnover for masses above  $10^{10.5} \Msun$ \citep[see e.g.,][]{Tremonti2004,Creasey2015,Segers2016}. This results from the efficiency of different feedback mechanisms: stellar feedback is the dominant effect in low-mass galaxies, while AGN feedback is the dominant effect for  $M_\ast{}>10^{10.5} \Msun$.

As a consequence of AGN feedback, galaxies above $10^{10.5} M_\odot$ host stars with a slightly lower metallicity than smaller galaxies, and thus host BBH mergers more efficiently. This in turn produces a higher number of merging BBHs in the local Universe, increasing the merger rate per galaxy, $n_{\rm GW}$.

\end{document}